\documentclass[aps,prx,reprint,superscriptaddress,amsmath,amssymb]{revtex4-2}

\usepackage{graphicx} % Include figure files
\usepackage{dcolumn}  % Align table columns on decimal point
\usepackage{bm}       % Bold math
\usepackage{hyperref} % Add hypertext capabilities
\usepackage{physics}  % Useful for braket notation and derivatives
\usepackage{xcolor}   % For colored text in drafts
\usepackage{tikz}     % Feynman-type diagrams in the Appendix

\providecommand{\per}{\operatorname{per}}
\providecommand{\hf}{\operatorname{hf}}
\providecommand{\E}{\mathbb{E}}

\newcommand{\Bset}{\mathcal{B}_{2k}}
\newcommand{\Pev}{\Pi^{\mathrm{ev}}_{2k}}
\newcommand{\cint}{\oint}

\begin{document}

\title{Moments of Random Permanents and Hafnians: A Replica Approach}

\author{Paolo Facchi}
\affiliation{Dipartimento Interateneo di Fisica, Università degli Studi di Bari Aldo Moro, I-70126 Bari, Italy}
\affiliation{INFN, Sezione di Bari, I-70125 Bari, Italy}

\author{Angelo Mariano}
\affiliation{ENEA, Italian National Agency for New Technologies, Energy and Sustainable Economic Development, I-70124 Bari, Italy}

\author{Giorgio Parisi}
\affiliation{Dipartimento di Fisica, Sapienza Università di Roma, I-00185 Rome, Italy}
\affiliation{INFN, Sezione di Roma 1, I-00185 Rome, Italy}
\affiliation{Institute of Nanotechnology (NANOTEC), CNR, Rome unit, I-00185 Rome, Italy}
\affiliation{International Research Center of Complexity Sciences, Hangzhou International Innovation Institute, Beihang University, Hangzhou 311115, China}

\author{Saverio Pascazio}
\affiliation{Dipartimento Interateneo di Fisica, Università degli Studi di Bari Aldo Moro, I-70126 Bari, Italy}
\affiliation{INFN, Sezione di Bari, I-70125 Bari, Italy}

\author{Antonello Scardicchio}
\affiliation{International Centre for Theoretical Physics (ICTP), I-34151 Trieste, Italy}
\affiliation{INFN, Sezione di Trieste, I-34128 Trieste, Italy}
\date{\today}

\begin{abstract}
Using contour-integral representations of the permanent and of the Hafnian, we turn the computation of their moments for random matrices into a replica calculation, analogous to that of the moments of the partition function of a mean-field spin glass. For the $2k$-th moment of $N\times N$ matrices with independent entries, the dominant saddle point is replica symmetric for $k>2$, and we obtain in closed form the exponential growth rate, the prefactor and the first $1/N$ correction, in agreement with exact enumerations. The growth rate depends only on whether the entries are real or complex; higher cumulants affect only the prefactor and the $1/N$ term. At $k=2$ the replica-symmetric solution becomes unstable, separating a power-law bulk of the distribution from a large-deviation tail, which we compute. The results apply to boson sampling with single photons and with squeezed light.\end{abstract}

\maketitle

\section{Introduction}

Determinants are famous for their cancellations; permanents for what
happens when those cancellations disappear. For an $N\times N$ matrix $M$,
the determinant
\[
\det M=\sum_{\sigma\in S_N}\operatorname{sgn}(\sigma)
\prod_i M_{i,\sigma(i)} 
\]
assigns a sign to each permutation.
Swapping two rows or two columns reverses its value, so identical rows or columns make it vanish.
This alternating structure permits efficient Gaussian elimination.

 In the eighteenth century,
Vandermonde studied products of differences that helped shape the theory
of alternating functions \cite{Vandermonde1774}, closely related to the determinant now bearing
his name~\cite{Ycart2013}. 
Cauchy subsequently spoke
of \emph{fonctions sym\'{e}triques permanentes} \cite{Cauchy1815}, and it was Muir ~\cite{Muir1882} who applied Cauchy's expression  ``permanent'' to the signless matrix sum
\[
\operatorname{per}M=\sum_{\sigma\in S_N}\prod_{i=1}^{N}M_{i,\sigma(i)}.
\]
The missing signs look like a minor edit, yet determinants admit
polynomial-time evaluation, while the inclusion--exclusion method
uses $O(N2^N)$ operations in the worst case~\cite{Ryser1963,Minc1978}. A remarkable difference in complexity.

Valiant proved that exact permanent evaluation is $\#\mathrm{P}$-complete
even for zero--one matrices~\cite{Valiant1979}. The class $\#\mathrm{P}$
concerns counting efficiently checkable solutions, and these zero--one
permanents count perfect matchings in bipartite graphs. This is a
worst-case exact result;
nonnegative permanents, for example, admit randomized relative
approximation~\cite{Jerrum2004}.

Caianiello translated the permutation bookkeeping into
physical terms. In seminal studies of quantum-field combinatorics, he
linked bosonic contractions to signless sums and related pairing
structures, placing permanents and Hafnians within the language of quantum particle statistics~\cite{Caianiello1953,Caianiello1959a,Caianiello1959b,Caianiello1973} \footnote{Hafnians were named by Caianiello ``to mark the fruitful period of stay in Copenhagen (Hafnia in Latin)'' \cite{Guerra2006}.}.
For noninteracting fermions, occupied one-particle
orbitals give a Slater determinant~\cite{Slater1929}. The difference between bosonic and fermionic exchange thus gives
the old distinction between two permutation sums a concrete physical setting.

Over the last two decades, this connection has become operational in photonics. Scheel showed in
2004 \cite{Scheel2004}, before the boson-sampling proposal of Aaronson and Arkhipov \cite{Aaronson2013},
that transition amplitudes of indistinguishable photons in passive
linear-optical networks involve permanents of interferometer
submatrices. For a given input-output photon pattern, the probability is proportional to the
squared modulus of such a permanent. The optical device generates samples
without returning the complex permanent: estimating the probability
of a very rare output by repeated counting can itself require many
trials. 

Sampling hardness and evaluating an individual permanent
are therefore distinct questions, as is the problem of establishing
that a measured distribution agrees with the intended one.
The proposed advantage concerns drawing such outputs; its strongest
claims for approximate, noisy sampling require additional complexity
conjectures~\cite{Aaronson2013}.

A number of mathematical problems remain open in this context. Tao and Vu focused on the permanents of random Bernoulli matrices, whose entries are independently chosen to be \(+1\) or \(-1\) with equal probability \cite{TaoVu2009}. While the determinant moments of matrices with independent Gaussian entries are well understood and can be computed exactly, the Bernoulli case is more intricate, and a systematic study of high-order moments appears to be lacking. In this article, we investigate the moments of permanents for both Gaussian and Bernoulli random matrices.

Integrated photonic experiments made the permanent connection
accessible to measurement. Broome and collaborators tested two- and
three-photon interference in a tunable six-mode circuit~\cite{Broome2013};
Spring and collaborators sampled three- and four-photon outputs on a
six-mode chip~\cite{Spring2013}. Tillmann and collaborators studied
three photons in five-mode written waveguides~\cite{Tillmann2013},
while Crespi and collaborators realized a laser-written five-mode
interferometer and compared three-photon counts with permanent-based
predictions~\cite{Crespi2013}. These experiments test the underlying
interference at small photon numbers; they do not themselves establish a
classical-computation speedup.

Validation studies subsequently tested
specific alternatives to genuine bosonic sampling~\cite{Spagnolo2014}.
In particular, Walschaers \emph{et al.} proposed efficiently accessible
statistics of two-mode photon-number correlations as a benchmark
against distinguishable-particle and selected mimic
models~\cite{Walschaers2016}; related signatures were later measured
on an integrated chip~\cite{Giordani2018}. 

\begin{figure}
    \includegraphics[width=0.95\columnwidth]{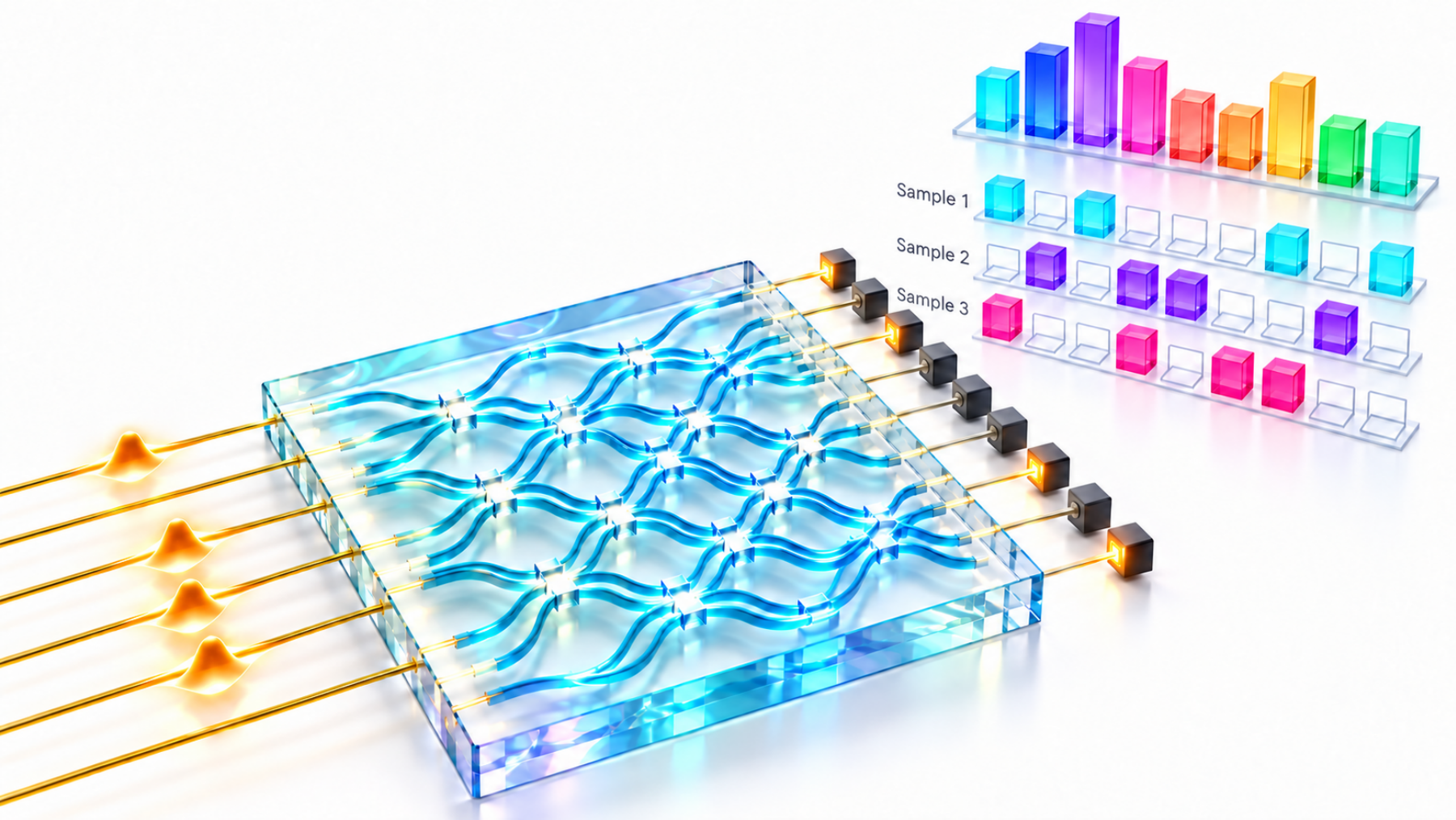}
    \caption{Scheme of a boson sampling experiment. Photons are injected into the input ports, interact with a linear system and are detected at the output ports. We shall assume that the system is lossless and no ``collisions'' occur (namely photons are detected at different output ports).}
    \label{fig:boson_sampling_nine_ports}
\end{figure}

A boson sampling experiment is deceptively simple: a certain number of photons are injected into the input ports, interact with a linear system, and ultimately are detected at the output ports. See the schematic in Fig. \ref{fig:boson_sampling_nine_ports}.
However, the underlying problem is extremely complex, involving advanced combinatorics, permanents, and Hafnians. 
The benchmark compares the
mean, standard deviation, and skewness of the correlations
across output-mode pairs, without reconstructing every output probability.
These selective tests do not
certify an arbitrary full output distribution. 

Another photonic route, \emph{Gaussian} boson sampling, employs squeezed inputs rather than fixed numbers of single photons. Hamilton
\emph{et al.} showed that its ideal, number-resolved output probabilities involve
\emph{Hafnians}, highlighting both the reach of photonic sampling and
the importance of distinguishing the underlying matrix
functions~\cite{Hamilton2017}.

In a recent large-scale experiment, Pan and collaborators realized
Gaussian boson sampling with 1,024 squeezed states in 8,176 modes,
registering events with up to 3,050 detector clicks and benchmarking
the outputs against classical models~\cite{Liu2026}. This samples
Gaussian outputs, rather than directly measuring individual permanents.
For random interferometers, even the
statistics of permanent-dependent outcomes pose subtle questions.
Random permanents have their own mathematical history~\cite {Borovskikh1994} and have not been systematically approached using the methods of statistical mechanics (a notable exception is \cite{Fyodorov2006}).

Here we study the statistics of random permanents and Hafnians, rather
than the evaluation of a prescribed instance. 

Section~II defines
normalized moments for real and circular complex random matrices and
derives the contour-integral representation of the permanent that will be used in the rest of the paper.

Section~III
turns the moments of real permanents into a replica calculation using
Hubbard-Stratonovich fields. For Gaussian entries, this yields an
auxiliary Hafnian representation; a large-$N$ saddle-point analysis shows the saddle point is replica symmetric for all moments with $k\geq 3$ and
provides the exponential growth rate, prefactor, and first $1/N$
correction. The section also shows how higher cumulants of symmetric
non-Gaussian entry distributions change the subleading terms without
changing the leading rate. 

Section~IV treats complex Gaussian
permanents, obtaining an exact moment--matrix-size duality and their
large-$N$ asymptotics. It also derives a duality between real Gaussian
permanents and Hafnians of symmetric complex Gaussian matrices, relevant
to Gaussian boson sampling.

In Section~V we use the moments to compute the full distribution of the normalized permanents, highlighting a transition region from the bulk (inaccessible to our methods) to the tail. In this transition region, a universal power law arises.

We compare numerical data for an ensemble of real matrices with the analytical calculation, and the agreement is remarkable. We remark that to access the distribution in the bulk, not just in the tail, one needs information about moments of order $k\leq 2$ which should require a different technique than the RS presented here, possibly an RSB calculation \cite{mezard1987spin,Parisi1979,Parisi1980a,Parisi1980b,Parisi2023}.

Section~VI makes the connection to photonic experiments explicit: in
the dilute, collision-free regime, single-photon boson-sampling
probabilities involve squared permanents of approximately Gaussian
interferometer submatrices, whereas squeezed-state Gaussian boson
sampling involves squared Hafnians. The moments describe the heavy
tails of ideal output probabilities, compare the two sampling settings,
and estimate the sample cost of cross-entropy and related statistical
benchmarks. These ensemble predictions do not by themselves certify an
individual device or establish the anti-concentration assumptions used
in sampling-hardness arguments. Section~\ref{sec:conclusions} summarizes the asymptotic
laws, their universality, and their limitations. 

The Appendix contains 
most of the technical derivations, including the saddle-point expansion,
phase zero modes, fluctuation determinants, and finite-size corrections.

We believe the new techniques we introduce in this paper could spur interest in the problem of permanents of random matrices, which has been for many years the less-explored counterpart to determinants in random matrix theory, largely due to the historical lack of tractable analytical tools. Determinants are, in fact, expressed in terms of the matrix spectrum, and there is a large literature treating them \cite{mehta2004random,livan2018introduction}. Moreover, our uncovering of replica symmetry and the possibility of replica symmetry breaking connects to another important subfield of statistical physics: spin glasses and disordered systems \cite{mezard1987spin}.

\section{Permanents, Hafnians and their integral representations}

We consider $N\times N$ random matrices $M$ whose elements are independent and identically distributed (i.i.d.) random variables with distribution $P(x)$. In general, $x$ can be real (in which case we require $P(x)$ to be even) or complex (in which case we require circular distributions, $P(e^{i\theta}x)=P(x)$). Such matrices are clearly not symmetric. 
We define the average $\langle\cdot\rangle$ as the average with respect to the probability measure $\prod_{ij}P(M_{ij})$.

For integer $k$ we aim to compute the moments of $\per M$
\begin{equation}
    \langle |\per M|^{2k}\rangle,\nonumber
\end{equation}
which we appropriately normalize to eliminate the factorials
\begin{equation}
  \label{eq:moment}
   \frac{\langle |\per M|^{2k}\rangle}{N!^{\,k}}\;=\;\langle Y^k\rangle  \;=\;\langle \exp(-k (-\ln(Y)))\rangle.
\end{equation}
In this way $\langle Y\rangle=1$: the moments may have only a possible exponential growth. We then define the rate function
\begin{equation}
  \Lambda_{\mathbb F}(k) \;=\; \lim_{N\to\infty}\frac1N \ln\langle Y^k\rangle,
  \label{eq:rate}
\end{equation}
with the subscript $\mathbb F$ equal to $\mathbb R$ for real and to $\mathbb C$ for complex matrices, which plays the role of a free energy.

Notice that we are evaluating 
\begin{equation}
  \int dM P(M) \exp \big(-k H \big),
\end{equation}
namely the partition function of a system with inverse temperature $k$ and Hamiltonian $H = - \ln (Y) \propto N$. We shall  
perform the transition from the canonical to the microcanonical ensemble. In the limit $N \to \infty$ we
will only need the saddle point approximation and the fluctuations around it.

One ensemble relevant for the boson sampling problem~\cite{AaronsonArkhipov2011} is that of {\it complex matrices}: the $N\times N$ submatrices of Haar-random $m\times m$ unitaries, rescaled by $\sqrt m$, are close to i.i.d.\ complex Gaussian matrices for $m$ large enough compared to $N$. We therefore take $M_{ij}$ i.i.d.\ circularly symmetric complex Gaussians, with 
distribution $P(z)=\frac{1}{\pi}e^{-|z|^2}$,
so that $\langle|M_{ij}|^2\rangle=1$ and $\langle M_{ij}^2\rangle=0$. 
Throughout the paper, the label $\mathbb R$ or $\mathbb C$ on $\Lambda$, on the prefactor $C_k$ and on the coefficient $a_1$ of the $1/N$ correction (see later) distinguishes real from complex entries.

In order to average over $M$, we use the representation of the permanent as a multiple integral in the complex plane which was introduced in \cite{Fyodorov2006}. For every $\rho>0$,
  \begin{equation}
    \label{eq:per-contour}
    \begin{split}
    \per (M) \;=\;& \prod_{i=1}^N\cint_{|u_i|=\rho}\frac{du_i}{2\pi i\,u_i^2}
    \prod_{j=1}^N\cint_{|v_j|=\rho}\frac{dv_j}{2\pi i\,v_j^2}\\
    &\times\exp\Big(\sum_{i,j=1}^N u_i\,M_{ij}\,v_j\Big).
    \end{split}
  \end{equation}

This representation is the twin of the very well-known representation of the determinant as an integral over Grassmann variables (as discussed in the Appendix)
\begin{equation}
    \det (M)=\int \prod_{i=1}^N d\psi_i d\overline{\psi}_i \exp\Big(\sum_{i,j=1}^N \overline{\psi}_i\,M_{ij}\,\psi_j\Big),
\end{equation}
which is widely used in the contexts of random matrix theory and Anderson localization \cite{efetov1999supersymmetry}.

The proof of Eq.\ \eqref{eq:per-contour} is straightforward: in the expansion of the entire function $e^{u^TMv}$, one should retain only the term $(u^TMv)^N$, and this contains all possible products $M_{i_1j_1}\cdots M_{i_Nj_N}$ where all the indices appear one and only one time, therefore making the sum over all the permutations (in the case of the determinant, reordering the Grassmann variables introduces the sign of the permutations).

For the Hafnian, we consider a $2N\times2N$ symmetric matrix $M$, $M_{ij}=M_{ji}$ and define
\begin{equation}
  \hf(M)\;=\;\sum_{T\in\mathcal T_{2N}}\ \prod_{\{i,j\}\in T}M_{ij},
\end{equation}
where $\mathcal T_{2N}$ is the set of the $(2N-1)!!$ perfect matchings (i.e., the partitions into pairs) $T$ of the set $[2N]=\{1,...,2N\}$.

The Hafnian has a contour representation analogous to~\eqref{eq:per-contour}, with a single set of variables, since rows and columns coincide:
\begin{equation}
  \label{eq:hf-contour}
  \hf(M)\;=\;\prod_{i=1}^{2N}\cint_{|u_i|=\rho}\frac{du_i}{2\pi i\,u_i^2}\;\exp\Big(\sum_{i<j}u_i\,M_{ij}\,u_j\Big).
\end{equation}
The proof is the same as for~\eqref{eq:per-contour}: in $(\sum_{i<j}M_{ij}u_iu_j)^N/N!$ the multilinear terms are the products of $N$ disjoint pairs covering $[2N]$, i.e.\ the perfect matchings, each one arising from $N!$ orderings. 

These representations, with the matrix in the exponent, turn the problem of a computing the moments of a permanent or a Hafnian into a form amenable to treatments using the full arsenal of statistical mechanics of disordered systems.

\section{Real matrices}

To compute $\langle|\per(M)|^{2k}\rangle=\langle(\per M)^{2k}\rangle$ we need now $2k\times N$ variables $u$ and $2k\times N$ variables $v$. We compactly write
\begin{equation}
     d\mu_{2k}(u,v) \;=\; \prod_{a=1}^{2k}\prod_{i=1}^N
  \frac{du_i^a}{2\pi i\,(u_i^a)^2}
  \prod_{a=1}^{2k}\prod_{j=1}^N\frac{dv_j^a}{2\pi i\,(v_j^a)^2}.
\end{equation}
With this notation
\begin{equation}
  \label{eq:cauchy}
  \begin{split}
  (\per M)^{2k}&=\cint d\mu_{2k}(u,v)\;\exp\Big(\sum_{i,j}M_{ij}X_{ij}\Big),\\
  X_{ij}&=\sum_{a=1}^{2k}u_i^a\,v_j^a.
  \end{split}
\end{equation}
We assume the distribution of $M_{ij}$ is even, $P(x)=P(-x)$, with unit variance ($\kappa_2=1$), and define as $\kappa_{m}$ its $m$-th cumulant 
\begin{equation}
    \psi(y)
    =\ln \langle e^{y X}\rangle =
    \ln\left(\int dx P(x)\,e^{y x}\right)=\sum_{m=1}^\infty\frac{\kappa_m}{m!} y^m.
\end{equation}
Cumulants can be seen also as connected correlation functions, in particular $\kappa_4=\kappa_2^2(K-3)$ where $K$ is the kurtosis. Prominent examples are Gaussian where $\kappa_2=1$ and $\kappa_{m\geq 3}=0$ and Rademacher ($X=\pm 1$) distribution where $\psi(y)=\ln(\cosh(y))$.

Now, by $\langle e^{M_{ij}X_{ij}}\rangle = e^{\psi (X_{ij})}$ we get
\begin{equation}
  \label{eq:averaged}
  \begin{split}
  \langle(\per M)^{2k}\rangle&= \cint d\mu_{2k}(u,v)\;\exp\Big(\sum_{i,j=1}^N\psi(X_{ij})\Big)\\
  &= \cint d\mu_{2k}(u,v)\;\exp\Big(\sum_{i,j=1}^N\sum_{m=1}^\infty\frac{\kappa_m}{m!}X_{ij}^m\Big).
  \end{split}
\end{equation}
The contour integral~\eqref{eq:cauchy} extracts the coefficient of the
monomial in which every $u_i^a$ and every $v_j^a$ has exponent one. Any
monomial divisible by some $(u_i^a)^2$ or $(v_j^a)^2$ therefore
integrates to zero, and so does any product containing such a monomial.
Consequently, if $R$ is a convergent power series all of whose
monomials are divisible by some square, then $e^{A+R}$ and $e^A$ have
the same integral. In
\begin{equation}
  \frac{X_{ij}^m}{m!} \;=\; \frac{1}{m!}\sum_{a_1,\dots,a_m}
  \prod_{l=1}^m u_i^{a_l}v_j^{a_l}
\end{equation}
the terms with a repeated replica index are of this kind, and each
subset $B$ of $m$ distinct replicas appears $m!$ times. Hence, under the
integral,
\begin{equation}
  \label{eq:reduction}
  \frac{X_{ij}^m}{m!}\;\longrightarrow\;
  \sum_{B\subseteq[2k],\ |B|=m} u_i^B\,v_j^B ,
\end{equation}
where $[2k]=\{1,...,2k\}$, which vanishes for $m>2k$, so that the series in~\eqref{eq:averaged}
becomes a finite sum. 
Summing over $m$,
\begin{eqnarray}
  \langle(\per M)^{2k}\rangle &=& \cint d\mu_{2k}(u,v)\;\exp\Big(\sum_{B\in\Bset}\kappa_{|B|}\,R_B\,S_B\Big),\nonumber\\
  R_B&=&\sum_{i=1}^N u_i^B,\qquad S_B=\sum_{j=1}^N v_j^B .
   \label{eq:overlap-coupling}
\end{eqnarray}
Here, 
\begin{equation}
    \Bset:=\bigl\{
        B\subseteq[2k]: |B|\in\{2,4,\ldots,2k\}
    \bigr\},
\end{equation}
denotes the family of nonempty even-cardinality subsets, and for every subset $B\subseteq[2k]$, we defined
\begin{equation}
    u_i^B:=\prod_{a\in B}u_i^a,
    \qquad
    v_j^B:=\prod_{a\in B}v_j^a.
\end{equation}
$R_B$ and $S_B$, with $B\subseteq[2k]$, are the row and column overlaps of the replicas in $B$:
for $|B|=2$, $R_{ab}=\sum_i u_i^au_i^b$ is the usual two-replica
overlap, and the blocks with $|B|\ge4$ are multi-replica overlaps,
weighted by the higher cumulants $\kappa_m$, with $m\geq4$. 

\subsection{Hubbard-Stratonovich decoupling}

We now introduce a complex Hubbard-Stratonovich (HS) integral. This is a standard trick to rewrite the average moments, and therefore the quenched average, of fully-connected statistical mechanics models like the SK model \cite{sherrington1975solvable}. In the context of Anderson localization (determinants) this leads to the nonlinear sigma model \cite{efetov1999supersymmetry} and in the context of random permanents it was used before in \cite{Fyodorov2006}. 

Introducing a new complex variable by defining an average
\begin{equation}
    \E_z(\cdot)=\int\frac{d^2z}{\pi}e^{-|z|^2}(\cdot)
\end{equation}
for an arbitrary number of complex variables $z$ indexed by a multi-index $B$, we have
\begin{equation}
    \label{eq:hs}
    \E_z\,e^{\kappa zR+\bar zS} \;=\; e^{\kappa RS},
\end{equation}
where $\kappa zR\equiv\sum_{B}\kappa_{|B|}z_{B}R_{B}$ and $\overline{z}S\equiv\sum_{B}\overline{z}_{B}S_{B},$ summing over all multi-indices $B\in\Bset$.

Using this,
\begin{equation}
  \label{eq:decoupled}
  \begin{split}
  \langle(\per M)^{2k}\rangle \;=\;& \E_z\;\cint d\mu_{2k}(u,v)\\
  &\times\exp\Big(\sum_{i=1}^N\sum_{B\in\Bset}\kappa_{|B|}\,z_B\,u_i^B\Big)\\
  &\times\exp\Big(\sum_{j=1}^N\sum_{B\in\Bset}\bar z_B\,v_j^B\Big).
  \end{split}
\end{equation}
In this way, the rows and columns have decoupled, as the sites do in the replica
calculation for spin glasses: the integrand is a product over $i$ of
functions of $(u_i^a)_a$ and over $j$ of functions of $(v_j^a)_a$, and
the contour measure factorizes in the same way.

Now let us go back to \eqref{eq:overlap-coupling} for the case $m=2$, which is the only term existing for a Gaussian distribution, we have explicitly $B=\{a,b\}$, $z_{B}=z_{ab}$, $R_B=R_{ab}$ and $S_B=S_{ab}$. One recognizes the spin glass overlaps $z_{ab}\sim Q_{ab}$. As usual, higher cumulants give rise to higher index overlaps $Q_{abcd}\sim z_{abcd}$ etc.

We can do the integrals over $d\mu_{2k}(u,v)$ analytically by~\eqref{eq:cauchy},
\begin{equation}
  \label{eq:single-row}
  \begin{split}
  &\prod_{a=1}^{2k}\cint_{|u^a|=\rho}\frac{du^a}{2\pi i\,(u^a)^2}\;\exp\Big(\sum_{B\in\Bset}c_B\,u^B\Big)\\
  &\qquad=\; \sum_{\pi\in\Pev}\ \prod_{B\in\pi}c_B ,
  \end{split}
\end{equation}
for any coefficients $c_B$, where $\Pev$ is the set of partitions of $[2k]$ into blocks of even size,
and $\pi\in\Pev$ is one such partition, a collection of disjoint subsets
$B\subseteq[2k]$ with $|B|$ even and $\ge2$ whose union is $[2k]$. 
Indeed, in
$\big(\sum_Bc_Bu^B\big)^n/n!$ a multilinear term in
$u^1,\dots,u^{2k}$ is a product of $n$ distinct blocks forming a
partition of $[2k]$, and each partition arises from $n!$ orderings.
Define
\begin{equation}
  \label{eq:F}
  F_\kappa(z) \;=\; \sum_{\pi\in\Pev}\ \prod_{B\in\pi}\kappa_{|B|}\,z_B,
  \qquad
  G(z) \;=\; \sum_{\pi\in\Pev}\ \prod_{B\in\pi}z_B,
\end{equation}
Each row contributes $F_\kappa(z)$ and each column contributes
$G(\bar z)=\overline{G(z)}$, the coefficients of $G$ being real.

Finally, we reach the HS form with one family of $z_B$'s, with $|B|=m$ for each cumulant $\kappa_m$ existing in the distribution, which is the central result of this paper:
\begin{equation}
    \label{eq:moment-general}
    \langle(\per M)^{2k}\rangle \;=\;
    \E_z\Big[F_\kappa(z)^N\ \overline{G(z)}^{\,N}\Big].
  \end{equation}

\subsection{The Gaussian distribution}

With this formal machinery, we will now solve the Gaussian case first. For Gaussian entries $\kappa_2=1$ and $\kappa_m=0$ for $m\ge4$, so every
partition containing a block of size $|B|\ge4$ has weight zero. The
partitions that survive have only blocks of size two, i.e.\ they are
the perfect matchings $T\in\mathcal{T}_{2k}$ of $[2k]$, and
\begin{equation}
  \label{eq:F-gauss}
  F_\kappa(z) \;=\; \sum_{T\in\mathcal{T}_{2k}}\ \prod_{\{a,b\}\in T}z_{ab}
  \;=\; \hf(z),
\end{equation}
the Hafnian of the HS fields matrix $(z_{ab})_{a<b}$. Also, we can keep in $\overline{G(z)}$ only the terms which are anti-holomorphic in $z_B$ with $|B|=2$ (the others being killed in the average $\E_z$) and it is not difficult to see that $G(z)=\hf(z)$ too. So for the real Gaussian case
\begin{equation}
    \label{eq:real-dual}
    \langle(\per M)^{2k}\rangle \;=\; \E_{z}|\hf(z)|^{2N},
\end{equation}
only over the variables $z_B$ with $B$ pairings of $[2k]$. Of these pairings, there are $D=k(2k-1)$. Changing to $z_B=y_B\sqrt{N}$ and using the homogeneity of the Hafnian, we have
\begin{equation}
    \langle(\per M)^{2k}\rangle \;=\;N^{kN+D}\int\prod_{e=1}^D
  \frac{d^2y_e}{\pi}\;
  \exp \left[-N\Phi_{\mathbb R}({\bm y})\right].
\end{equation}
where 
\begin{equation}
   \Phi_{\mathbb R}=\sum_{e=1}^D|y_e|^2-\ln\hf({\bm y})-\ln\overline{\hf({\bm y})}. 
   \label{eq:action}
\end{equation}

We can now compute the integral using the saddle point method. For large $N$ we find the stationarity condition, $\partial_{\overline{y}_e}\Phi_{\mathbb R}=0$, gives
\begin{equation}
  \label{eq:field-saddle}
  y_e \;=\; \overline{\left(\frac{\partial_e\hf(y)}{\hf(y)}\right)}.
\end{equation}
Multiplying by $\overline{y}_e$, summing over $e=1,...,D$, and using homogeneity
$\sum_{e=1}^D y_e\,\partial_e\hf=k\hf$ gives $\sum_{e=1}^D|y_e|^2=k$ at any saddle.

There is a replica-symmetric (RS) saddle $y_e=y_0$ for all $e=1,...,D$. One can prove, using inequalities in \cite{roos2025new} that indeed, for $k\geq3$, the RS saddle is a global minimum of the action~\eqref{eq:action}. It is then not difficult to get that $\partial_e\hf=(2k-3)!!\,y_0^{k-1}$ and 
$\hf=(2k-1)!!\,y_0^{k}$, so the
saddle equation~\eqref{eq:field-saddle} reads $y_0=1/\bigl((2k-1)y_0\bigr)$:
\begin{equation}
  \label{eq:field-y0}
  y_0^2 \;=\; \frac{1}{2k-1} \;=\; q_0 .
\end{equation}
The saddle value of $|y_e|^2$ is the RS overlap. In general
$|y_e|^2=q_e$ at a saddle, because $|z_e|^2\approx Nq_e$: the field $z_e$
is conjugate to the number of rows in which $a$ and $b$ are paired. The
exponent at the saddle $\bm{y}_s=(y_0,...,y_0)$ gives the leading order exponential in $N$. The rate function (\ref{eq:rate})
\begin{equation}
  \label{eq:realLambda}
  \begin{split}
  \Lambda_{\mathbb R}(k)&=k-\Phi_{\mathbb R}(\bm{y}_s)\\
  &=2k\ln2 + 2\ln\Gamma\big(k+\tfrac12\big) - \ln\pi - k\ln(2k-1),
  \end{split}
\end{equation}
it is an analytic function of $k>2$, and the solution of the problem at leading exponential order in $N$:
\begin{equation}
    \lim_{N\to\infty}\frac{1}{N}\ln{\langle Y^k\rangle}=\Lambda_{\mathbb R}(k).\nonumber
\end{equation}

The polynomial prefactor and the $1/N$ correction come from Gaussian fluctuations around the saddle point (which has flat directions) and from the next order in the saddle-point expansion. The calculation is detailed in the Appendix, and its result is
\begin{equation}
  \label{eq:field-result}
  \langle Y^k\rangle \;=\; C^{\mathbb R}_k\Bigl(1+\frac{a^{\mathbb R}_1}{N}+O\Bigl(\frac{1}{N^2}\Bigr)\Bigr)\,N^{k/2}e^{N\Lambda_{\mathbb R}(k)},
\end{equation}
where
\begin{equation}
  \label{eq:Ck}
  \begin{split}
  C^{\mathbb R}_k\;=\;& 2^{k-1}(2\pi)^{k/2}\Big(\frac{k-1}{2k-1}\Big)^{k-\frac12}\\
  &\times\Big(1+\frac{1}{4(k-1)(k-2)}\Big)^{k(2k-3)/2},
  \end{split}
\end{equation}
and
\begin{equation}
  \label{eq:a1}
  a^{\mathbb R}_1(k) \;=\; \frac{k\,\big(8k^4-44k^3+124k^2-178k+93\big)}
                    {48\,(k-1)\,(k-2)^2\,(2k-5)} .
\end{equation}
For large $k$, $a^{\mathbb R}_1\sim k/12$ so the saddle point approximation is valid for $k>2$ and $N\gg k/12$. Surprisingly, at $k=3, N=9$, the error relative to the numerically exact values goes from $\sim 29\%$ at leading order to about $3\%$ with the $1/N$ corrections included.

\begin{table*}
\caption{Moments $\langle Y^k\rangle=\langle(\per M)^{2k}\rangle/N!^{\,k}$
for real Gaussian entries: exact values from~\eqref{eq:real-dual}, leading saddle-point prediction
$C^{\mathbb R}_k\,N^{k/2}e^{N\Lambda_{\mathbb R}(k)}$ from~\eqref{eq:field-result}, and the same including the factor $1+a^{\mathbb R}_1/N$. Relative deviations from the exact values in parentheses.}
\label{tab:real}
\begin{ruledtabular}
\begin{tabular}{cclll}
$N$ & $k$ & exact & leading & with $1/N$ \\
\hline
7  & 3 & $1.74170\times10^{4}$   & $1.22806\times10^{4}$ ($-29.5\%$)  & $1.96819\times10^{4}$ ($+13.0\%$) \\
9  & 3 & $8.23966\times10^{4}$   & $5.80073\times10^{4}$ ($-29.6\%$)  & $8.51982\times10^{4}$ ($+3.4\%$) \\
11 & 3 & $3.52599\times10^{5}$   & $2.53953\times10^{5}$ ($-28.0\%$)  & $3.51349\times10^{5}$ ($-0.4\%$) \\
13 & 3 & $1.42468\times10^{6}$   & $1.05712\times10^{6}$ ($-25.8\%$)  & $1.40017\times10^{6}$ ($-1.7\%$) \\
\hline
2  & 4 & $2520.00$                & $2064.67$ ($-18.1\%$)               & $3491.31$ ($+38.5\%$) \\
3  & 4 & $3.10800\times10^{4}$   & $2.13315\times10^{4}$ ($-31.4\%$)  & $3.11577\times10^{4}$ ($+0.3\%$) \\
\end{tabular}
\end{ruledtabular}
\end{table*}

\subsection{General distributions}
We start this section by briefly showing that $\Lambda_{\mathbb R}(k)$ is, in fact, the same for all distributions $P(x)$ with $x$ real which possess all moments. We do this by showing that, in the saddle point approximation, the corrections to the formula of the average moments $\langle Y^k\rangle$, appear only in $C^{\mathbb R}_k$, which gets corrected by a factor $e^{W_k}$ proportional to the fourth cumulant $\kappa_4$, while the sixth and higher cumulants contribute only at order $1/N$ (which we also compute).

The idea is that at the RS saddle only the pair fields are macroscopic, $|z_e|^2\approx Nq_0$, while the fields $z_B$ with $|B|\ge4$, conjugate to the multi-replica overlaps, remain of order one. We therefore rescale only the former, $z_e=\sqrt N y_e$ for $e=1,\dots,D$. A partition $\pi\in\Pev$ containing $p$ pairs then contributes to $F_\kappa$ with a power $N^{p/2}$: the pairings, $p=k$, give back the Hafnian, a partition with one block of size four has $p=k-2$ and is suppressed by $1/N$, and those with a block of size six or two blocks of size four by $N^{-3/2}$ or more. Hence
\begin{equation}
  \begin{split}
  F_\kappa(z)\;=\;N^{k/2}\,\hf({\bm y})\Big[1&+\frac{\kappa_4}{N}\sum_{B\subseteq[2k],\ |B|=4} z_B\,\frac{h_B(\bm{y})}{\hf(\bm{y})}\\
  &+O(N^{-3/2})\Big],
  \end{split}
\end{equation}
where $h_B({\bm y})$ is the Hafnian of the pair fields $y_{ab}$ with $a,b\in[2k]\setminus B$. The same expansion holds for $\overline{G(z)}$, with $\kappa_4$ replaced by $1$ and $z_B$ by $\bar z_B$. Raising to the power $N$, the four-replica fields appear linearly in the exponent with coefficients of order one, and the average over each of them is done with~\eqref{eq:hs}:
\begin{equation}
  \begin{split}
  \langle(\per M)^{2k}\rangle \;=\;& N^{kN+D}\int\prod_{e=1}^D\frac{d^2y_e}{\pi}\;e^{-N\Phi_{\mathbb R}({\bm y})}\\
  &\times\exp\Big(\kappa_4\sum_{B\subseteq[2k],\ |B|=4}\Big|\frac{h_B({\bm y})}{\hf({\bm y})}\Big|^2\Big)\\
  &\times\big[1+O(1/N)\big].
  \end{split}
\end{equation}
The new factor is of order one, so it does not move the saddle point and $\Lambda_{\mathbb R}(k)$ is unchanged. 

This is the proof that the rate $\Lambda_{\mathbb R}(k)$ is universally valid for all real matrices whose entries' distribution possesses all the cumulants. At the RS saddle $h_B(\bm{y}_s)=(2k-5)!!\,y_0^{k-2}$, so that
\begin{equation}
    \frac{h_B(\bm{y}_s)}{\hf(\bm{y}_s)}=\frac{1}{(2k-1)(2k-3)y_0^2}=\frac{1}{2k-3}
\end{equation}
for each of the $\binom{2k}{4}$ blocks, and the prefactor becomes $C^{\mathbb R}_k\to e^{W_k}C^{\mathbb R}_k$ with
\begin{equation}
  \label{eq:Wk}
  W_k\;=\;\kappa_4\binom{2k}{4}\frac{1}{(2k-3)^2}\;=\;\kappa_4\,\frac{k(k-1)(2k-1)}{6(2k-3)} .
\end{equation}
Taking as an example Rademacher $M_{ij}=\pm1$ entries, we have $\kappa_2=1$, $\kappa_4=-2$, and $\kappa_6=16$. Therefore,
\begin{equation}
  W_k^{\pm 1}\;=\;-\frac{k(k-1)(2k-1)}{3(2k-3)},
\end{equation}
while the large-$N$ prefactor $e^{W_k}C^{\mathbb R}_k$ takes the values $0.386$, $0.0905$, $0.0112$ for $k=3,4,5$. For $k=2$ the same formula gives $e^{W_2}=e^{-2}$, which is the exact ratio of the fourth moments of the $\pm1$ and Gaussian permanents at large $N$. The terms neglected above, the six-replica fields weighted by $\kappa_6=16$ and the products of two four-replica fields, only change the coefficient of the $1/N$ correction, so for $\pm1$ entries the $1/N$ coefficient differs from the Gaussian value~\eqref{eq:a1}. 

Dividing $\langle(\per M)^{2k}\rangle$ by $N!^k$ we obtain $\langle Y^k\rangle$ and, using Stirling, we conclude with the formula for a general symmetric law correct to $O(1/N)$:
\begin{equation}
  \begin{split}
  \langle Y^k\rangle \;=\;& e^{W_k}\,C^{\mathbb R}_k\,N^{k/2}e^{N\Lambda_{\mathbb R}(k)}\\
  &\times\Big[1+\frac{a^{\mathbb R}_1(k)+w_1(k)}{N}+O(N^{-2})\Big],\\
  w_1(k)\;=\;&\alpha(k)\,\kappa_4+\beta(k)\,\kappa_4^2+\gamma(k)\,\kappa_6 ,
  \end{split}
\end{equation}
with $a^{\mathbb R}_1$ defined in~\eqref{eq:a1} and 
\begin{equation}
  \begin{split}
  \alpha&=\frac{k(2k-1)^2}{6(2k-3)(2k-5)},\\
  \beta&=-\frac{k(k-1)(k-2)(2k-1)^2}{18(2k-3)^2},\\
  \gamma&=\frac{k(k-1)(k-2)(2k-1)^2}{90(2k-3)(2k-5)} ,
  \end{split}
\end{equation}
while $\kappa_8$ and higher cumulants enter only at order $N^{-2}$. The accuracy of this formula is shown in Table~\ref{tab:pm1}.

\begin{figure}
\includegraphics[width=\columnwidth]{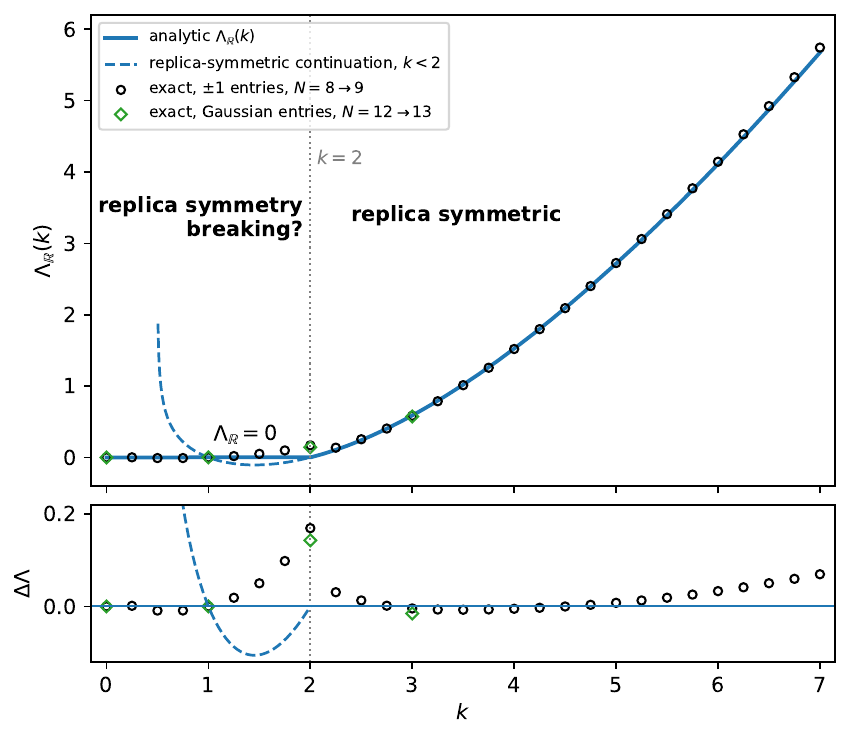}
\caption{Rate function $\Lambda_{\mathbb R}(k)$. Solid line: $\Lambda_{\mathbb R}=0$ for $k\le2$ and the replica-symmetric result~\eqref{eq:realLambda} for $k>2$. Dashed line: continuation of~\eqref{eq:realLambda} to $k<2$. Symbols: $\ln\big(\langle Y^k\rangle_{N+1}/\langle Y^k\rangle_N\big)-\frac k2\ln\frac{N+1}{N}$, the last term only for $k>2$, from the exact enumeration of $\pm1$ matrices with $N=8,9$ (circles) and from the exact moments of real Gaussian matrices with $N=12,13$ (diamonds). Lower panel: difference from the solid line.}
\label{fig:LambdaR}
\end{figure}

The comparison of the numerical data with the analytic formulas is presented in Fig.~\ref{fig:LambdaR}.

\begin{table*}
\caption{Moments $\langle Y^k\rangle=\langle(\per M)^{2k}\rangle/N!^{\,k}$
for $\pm1$ entries: exact values (from complete enumeration, $2^{49}$ matrices for $N=8$ and $2^{64}$ for $N=9$), leading saddle-point prediction
$e^{W_k}C^{\mathbb R}_k\,N^{k/2}e^{N\Lambda_{\mathbb R}(k)}$, and the same including the factor
$1+(a^{\mathbb R}_1+w_1)/N$. Relative deviations from the exact values in parentheses.}
\label{tab:pm1}
\begin{ruledtabular}
\begin{tabular}{cclll}
$N$ & $k$ & exact & leading & with $1/N$ \\
\hline
8 & 3 & $1059.70$              & $963.46$ ($-9.1\%$)              & $1092.39$ ($+3.1\%$) \\
8 & 4 & $1.21105\times10^{6}$  & $1.14509\times10^{6}$ ($-5.4\%$) & $1.21821\times10^{6}$ ($+0.6\%$) \\
8 & 5 & $5.3046\times10^{9}$   & $5.5446\times10^{9}$ ($+4.5\%$)  & $5.4275\times10^{9}$ ($+2.3\%$) \\
\hline
9 & 3 & $2265.88$              & $2069.35$ ($-8.7\%$)             & $2315.51$ ($+2.2\%$) \\
9 & 4 & $7.00192\times10^{6}$  & $6.65476\times10^{6}$ ($-5.0\%$) & $7.03248\times10^{6}$ ($+0.4\%$) \\
9 & 5 & $1.08528\times10^{11}$ & $1.12564\times10^{11}$ ($+3.7\%$)& $1.10452\times10^{11}$ ($+1.8\%$) \\
\end{tabular}
\end{ruledtabular}
\end{table*}

\section{Complex Gaussian matrices}

In this section we compute the moments of permanents and of Hafnians for the complex, circular ensemble of the $M_{ij}$.

\subsection{Permanent}

The integral representation has the same $2k\times N$ variables $u_i^a$, $v_j^a$ and the same measure $d\mu_{2k}(u,v)$ as in~\eqref{eq:cauchy}, but we must split the replicas into two groups: $a\in[k]$ for the $k$ copies of $\per M$, and $b\in[2k]\setminus[k]=\{k+1,\dots,2k\}$ for the $k$ copies of $\overline{\per M}$. Throughout this section $a$ runs over $[k]$ and $b$ over $[2k]\setminus[k]$. Using~\eqref{eq:per-contour} for each copy,
\begin{equation}
  \label{eq:cauchy-c}
  \begin{split}
  |\per M|^{2k}&=\cint d\mu_{2k}(u,v)\\
  &\quad\times\exp\Big(\sum_{i,j=1}^N\big[M_{ij}X^+_{ij}+\overline{M}_{ij}X^-_{ij}\big]\Big),\\
  X^+_{ij}&=\sum_{a=1}^{k}u_i^a\,v_j^a,\qquad X^-_{ij}=\sum_{b=k+1}^{2k}u_i^b\,v_j^b ,
  \end{split}
\end{equation}
so that $X^+_{ij}+X^-_{ij}$ is the $X_{ij}$ of~\eqref{eq:cauchy}. The variables with $b>k$ are independent integration variables, not the complex conjugates of those with $a\le k$. The complex Gaussian generating function is exactly bilinear, $\langle e^{xM_{ij}+y\overline{M}_{ij}}\rangle=e^{xy}$ for any complex $x,y$, so that, in place of~\eqref{eq:averaged},
\begin{equation}
  \label{eq:overlap-c}
  \begin{split}
  \langle|\per M|^{2k}\rangle&=\cint d\mu_{2k}(u,v)\;\exp\Big(\sum_{i,j=1}^N X^+_{ij}X^-_{ij}\Big)\\
  &=\cint d\mu_{2k}(u,v)\;\exp\Big(\sum_{a=1}^{k}\sum_{b=k+1}^{2k}R_{ab}\,S_{ab}\Big),
  \end{split}
\end{equation}
with $R_{ab}=\sum_iu_i^au_i^b$ and $S_{ab}=\sum_jv_j^av_j^b$ the overlaps of~\eqref{eq:overlap-coupling}. Two differences with the real case are worth noticing. First, we don't need the reduction~\eqref{eq:reduction}, since there are no higher powers of $X$ to discard. Second, only the $k^2$ overlaps between a replica $a\in[k]$ and a replica $b\in[2k]\setminus[k]$ appear, instead of all the $D=k(2k-1)$ pairs of $[2k]$: since $\langle M_{ij}^2\rangle=\langle\overline{M}_{ij}^2\rangle=0$, two copies of $\per M$, or two copies of $\overline{\per M}$, are never contracted. The replica graph is the complete bipartite graph $K_{k,k}$ rather than the complete graph on $[2k]$.

We decouple each of the $k^2$ terms with~\eqref{eq:hs} at $\kappa=1$, introducing the fields $z_{ab}$. Rows and columns decouple as in~\eqref{eq:decoupled}, and the single-row integral is~\eqref{eq:single-row} with $c_B=z_{ab}$ for $B=\{a,b\}$ and $c_B=0$ for all other blocks. The only surviving partitions are the perfect matchings of $[2k]$ in which every $a\in[k]$ is paired with some $b\in[2k]\setminus[k]$, i.e.\ the bijections $\sigma:[k]\to[2k]\setminus[k]$, so that each row contributes
\begin{equation}
  \label{eq:F-complex}
  \sum_{\sigma}\prod_{a=1}^{k}z_{a,\sigma(a)}\;=\;\per(z),
\end{equation}
the permanent of the $k\times k$ matrix $(z_{ab})$ with rows $a\in[k]$ and columns $b\in[2k]\setminus[k]$. Equivalently, $\per(z)$ is the Hafnian~\eqref{eq:F-gauss} of the field matrix with $z_{aa'}=z_{bb'}=0$. Each column contributes $\per(\bar z)=\overline{\per(z)}$, and
\begin{equation}
  \label{eq:moment-complex}
  \langle|\per M|^{2k}\rangle\;=\;\E_z\,|\per(z)|^{2N},
\end{equation}

All the integrals are absolutely convergent, and the contours are compact, so~\eqref{eq:moment-complex} is an exact identity. Since $\E_z$ is the average over a $k\times k$ matrix of i.i.d.\ circular complex Gaussians with unit variance, \eqref{eq:moment-complex} is a self-duality between the size of the matrix and the order of the moment,
\begin{equation}
  \label{eq:duality}
  \langle|\per M|^{2k}\rangle_{N}\;=\;\langle|\per M|^{2N}\rangle_{k},
\end{equation}
where the subscript denotes the size of $M$.

This moment-size duality can be seen as a special case of Theorem 6.2 in \cite{Fyodorov2006}.\footnote{We thank Yan Fyodorov for pointing this out to us.} was obtained in~\cite{Nezami2021} using representation theory. Notice that such a duality exists also for the SK model \cite{sherrington1975solvable} in which the number of replicas can be exchanged with the number of spins in the partition function. 

The real matrix case instead has no such symmetry and the dual object is a Hafnian. For $k=1$, \eqref{eq:moment-complex} gives $\E_z|z|^{2N}=N!$. For $k=2$, $\per(z)=z_{13}z_{24}+z_{14}z_{23}$ is the sum of two independent phase-invariant terms, so that
\begin{equation}
  \label{eq:k2-complex}
  \begin{split}
  \langle|\per M|^{4}\rangle&=\sum_{n=0}^{N}\binom{N}{n}^2(n!)^2\,\big((N-n)!\big)^2\\
  &=(N+1)\,N!^{\,2},
  \end{split}
\end{equation}
i.e.\ $\langle Y^2\rangle=N+1$ exactly, for every $N$.

Rescaling $z_{ab}=\sqrt N\,y_{ab}$ and using the homogeneity of the permanent,
\begin{equation}
  \begin{split}
  \langle|\per M|^{2k}\rangle\;=\;&N^{kN+k^2}\int\prod_{a=1}^{k}\prod_{b=k+1}^{2k}\frac{d^2y_{ab}}{\pi}\\
  &\times\exp\left[-N\Phi_{\mathbb C}({\bm y})\right],
  \end{split}
\end{equation}
where $\Phi_{\mathbb C}=\sum_{a,b}|y_{ab}|^2-\ln\per({\bm y})-\ln\overline{\per({\bm y})}$. For fixed $k$ this is an integral over $2k^2$ real variables with large parameter $N$. The stationarity condition $\partial_{\overline{y}_{ab}}\Phi_{\mathbb C}=0$ reads
\begin{equation}
  y_{ab}\;=\;\overline{\left(\frac{\partial_{ab}\per({\bm y})}{\per({\bm y})}\right)},
\end{equation}
and homogeneity gives again $\sum_{a,b}|y_{ab}|^2=k$ at any saddle. At the replica-symmetric saddle $y_{ab}=y_0$ one has $\per=k!\,y_0^{k}$ and $\partial_{ab}\per=(k-1)!\,y_0^{k-1}$, so that $y_0=1/(ky_0)$:
\begin{equation}
  \label{eq:y0-complex}
  y_0^2\;=\;\frac1k\;=\;q_0^{\mathbb C}.
\end{equation}
Row by row, each replica $a\in[k]$ is paired with one of the $k$ replicas $b\in[2k]\setminus[k]$, instead of one of the $2k-1$ other replicas as in~\eqref{eq:field-y0}. Dividing by $N!^{\,k}$, the rate function~\eqref{eq:rate} is
\begin{equation}
  \label{eq:rate-complex}
  \Lambda_{\mathbb C}(k)\;=\;k-\Phi_{\mathbb C}(\bm{y}_s)\;=\;2\ln\Gamma(k+1)-k\ln k .
\end{equation}

Both $\Lambda^{\mathbb{R}}$ and $\Lambda^{\mathbb{C}}$ have the form $2\ln(\text{number of perfect matchings of the replica graph})-k\ln(\text{degree of the replica graph})$, with $(2k-1)!!$ and $2k-1$ for the complete graph on $[2k]$, and $k!$ and $k$ for $K_{k,k}$. This is probably due to their representation-theoretic meaning, which is how they were computed in \cite{Nezami2021}.

One has $\Lambda_{\mathbb C}(1)=\Lambda_{\mathbb C}(2)=0$ and $\Lambda_{\mathbb C}(3)=\ln\frac43\simeq0.288$, against $\Lambda_{\mathbb R}(3)=\ln\frac95\simeq0.588$ for real Gaussian entries: the right tail of $|\per M|^2$ is lighter for complex entries.

Integrating the Gaussian fluctuations around the saddle point and over its flat directions, and including the next order of the saddle-point expansion (see the Appendix), we find
\begin{equation}
  \label{eq:result-complex}
  \langle Y^k\rangle\;=\;C^{\mathbb C}_k\,\big(1+a^{\mathbb C}_1/N+O(1/N^2)\big)\,N^{(k-1)/2}e^{N\Lambda_{\mathbb C}(k)},
\end{equation}
where
\begin{equation}
  \label{eq:Ck-complex}
  C^{\mathbb C}_k\;=\;(2\pi)^{\frac{k-1}{2}}\,k^{-\frac12}\Big(\frac{(k-1)^2}{k(k-2)}\Big)^{(k-1)^2/2},
\end{equation}
and
\begin{equation}
  \label{eq:a1-complex}
  a^{\mathbb C}_1(k)\;=\;\frac{(k-1)\big(k^3-5k^2+14k-12\big)}{12\,(k-2)^3} .
\end{equation}
The replica-symmetric saddle is the global maximum of $|\per({\bm y})|$ on the sphere $\sum_{a,b}|y_{ab}|^2=k$, as we checked numerically for $k=3,4,5$; this should follow from a Hadamard-type inequality for permanents~\cite{CarlenLiebLoss2006}. Since for fixed $k$ the integral is finite dimensional, \eqref{eq:result-complex} then gives the exact asymptotics.

For $k=2$ one of the eigenvalues of the Hessian at the saddle point, the coefficient of $(\Re\zeta_{ab})^2$ in Eq.~\eqref{eq:hessian-complex} of the Appendix, vanishes, which is why $C^{\mathbb C}_k$ and $a^{\mathbb C}_1$ are singular there: the $(k-1)^2=1$ additional flat direction gives one more power of $N^{1/2}$, in agreement with the exact $\langle Y^2\rangle=N+1$ of~\eqref{eq:k2-complex}. For $k<2$ this eigenvalue is negative, and the replica-symmetric saddle is unstable. This is consistent with the fact that $\Lambda_{\mathbb C}$, being convex in $k$ and vanishing at $k=0,1,2$, vanishes identically on $[0,2]$: the expression~\eqref{eq:rate-complex} holds for $k\ge2$, and its continuation to $0<k<2$ does not give the rate function.

In~\cite{Nezami2021} the moments are normalized by $\mathcal N_{N,k}=N!^{\,2k}k!^{\,2N}/(kN)!$, which is a lower bound to $\langle|\per M|^{2k}\rangle$ and, like it, is invariant under $N\leftrightarrow k$. Stirling's formula gives $(kN)!=k^{kN}N!^{\,k}\,k^{\frac12}(2\pi N)^{-\frac{k-1}{2}}\big(1+O(1/N)\big)$, so that
\begin{equation}
  \begin{split}
  \mathcal N_{N,k}&=N!^{\,k}\,\Big(\frac{k!^{\,2}}{k^k}\Big)^{N}(2\pi N)^{\frac{k-1}{2}}k^{-\frac12}\big(1+O(1/N)\big)\\
  &=N!^{\,k}\,(2\pi)^{\frac{k-1}{2}}k^{-\frac12}N^{\frac{k-1}{2}}e^{N\Lambda_{\mathbb C}(k)}\big(1+O(1/N)\big),
  \end{split}
\end{equation}
where we used~\eqref{eq:rate-complex}. The normalization $\mathcal N_{N,k}$ has therefore the same rate and the same power of $N$ as~\eqref{eq:result-complex}, which is equivalent to
\begin{equation}
  \label{eq:nezami-limit}
  \lim_{N\to\infty}\frac{\langle|\per M|^{2k}\rangle}{\mathcal N_{N,k}}\;=\;\Big(\frac{(k-1)^2}{k(k-2)}\Big)^{(k-1)^2/2},
\end{equation}
equal to $16/9$, $(9/8)^{9/2}\simeq1.699$ and $(16/15)^8\simeq1.676$ for $k=3,4,5$, and decreasing to $\sqrt e\simeq1.649$ for $k\to\infty$. By~\eqref{eq:duality} this is also the limit of large moments of the permanent of $k\times k$ matrices. The existence of this limit, with a value no larger than $2$, was conjectured in~\cite{Nezami2021} based on exact moments and a representation-theoretic lower bound equal to $13/8$ for $k\ge4$. Incidentally, $13/8=1+\frac12+\frac18$ used by Nezami in \cite{Nezami2021} is the truncation at second order of $\sqrt e=\sum_{n\ge0}2^{-n}/n!$.

In Table~\ref{tab:complex} we compare~\eqref{eq:result-complex} with exact moments. These are obtained by evaluating~\eqref{eq:moment-complex} as a finite sum: expanding $\per(z)^N$ and using $\E_z|\prod_{a,b}z_{ab}^{A_{ab}}|^2=\prod_{a,b}A_{ab}!$, one gets $\langle|\per M|^{2k}\rangle=\sum_A c_N(A)^2\prod_{a,b}A_{ab}!$, where $A$ runs over the $k\times k$ matrices of non-negative integers with all row and column sums equal to $N$, and $c_N(A)$ is the number of ordered $N$-tuples of permutations of $[k]$ whose permutation matrices sum to $A$. For $k=4$ the values agree with those of~\cite{Nezami2021}. Including the factor $1+a^{\mathbb C}_1/N$, the deviation at $k=3$ drops from $-3.3\%$ to $-0.1\%$ at $N=60$. For $k=4$ exact moments are available only up to $N=10$, and the residual deviation of about $3\%$ is presumably due to the higher orders in $1/N$.

\begin{table*}
\caption{Moments $\langle Y^k\rangle=\langle|\per M|^{2k}\rangle/N!^{\,k}$
for complex Gaussian entries: exact values were obtained from~\eqref{eq:moment-complex}, and leading saddle-point prediction
$C^{\mathbb C}_k\,N^{(k-1)/2}e^{N\Lambda_{\mathbb C}(k)}$ from~\eqref{eq:result-complex}, and the same including the factor $1+a^{\mathbb C}_1/N$. Relative deviations from the exact values in parentheses.}
\label{tab:complex}
\begin{ruledtabular}
\begin{tabular}{cclll}
$N$ & $k$ & exact & leading & with $1/N$ \\
\hline
10 & 3 & $1307.17$               & $1145.21$ ($-12.4\%$)              & $1374.25$ ($+5.1\%$) \\
20 & 3 & $4.50192\times10^{4}$   & $4.06726\times10^{4}$ ($-9.7\%$)   & $4.47398\times10^{4}$ ($-0.6\%$) \\
40 & 3 & $2.69984\times10^{7}$   & $2.56511\times10^{7}$ ($-5.0\%$)   & $2.69337\times10^{7}$ ($-0.2\%$) \\
60 & 3 & $1.25502\times10^{10}$  & $1.21331\times10^{10}$ ($-3.3\%$)  & $1.25375\times10^{10}$ ($-0.1\%$) \\
\hline
6  & 4 & $2.97430\times10^{4}$   & $2.55122\times10^{4}$ ($-14.2\%$)  & $2.92327\times10^{4}$ ($-1.7\%$) \\
8  & 4 & $2.28036\times10^{5}$   & $1.98848\times10^{5}$ ($-12.8\%$)  & $2.20597\times10^{5}$ ($-3.3\%$) \\
10 & 4 & $1.57675\times10^{6}$   & $1.40686\times10^{6}$ ($-10.8\%$)  & $1.52996\times10^{6}$ ($-3.0\%$) \\
\end{tabular}
\end{ruledtabular}
\end{table*}

Finally, for a general circularly symmetric law, $P(M)=P(e^{i\theta}M)$, the generating function depends on $xy$ only, $\ln\langle e^{xM+y\overline{M}}\rangle=\sum_{m\ge1}\kappa^{\mathbb C}_m\,(xy)^m/(m!)^2$ with $\kappa^{\mathbb C}_1=\langle|M|^2\rangle=1$ (for unit-modulus entries $\kappa^{\mathbb C}_2=-1$). Applying~\eqref{eq:reduction} to $X^+$ and $X^-$ separately, the cumulant $\kappa^{\mathbb C}_m$ couples a block of $m$ replicas of $[k]$ with a block of $m$ replicas of $[2k]\setminus[k]$, and the analogue of~\eqref{eq:moment-general} holds with partitions of $[2k]$ into such balanced blocks. The power counting of the previous section applies unchanged, so that $\Lambda_{\mathbb C}(k)$ is universal within this class.

\subsection{Hafnian}

In Gaussian boson sampling the single photons at the input are replaced by single-mode squeezed states~\cite{Hamilton2017}.
The probability of detecting one photon in each mode of a set $S$ of $2N$ output modes is then $p_S=|\hf(B_S)|^2/\prod_j\cosh r_j$, where $r_j$ are the squeezing parameters, $B=U\,\mathrm{diag}(\tanh r_j)\,U^T$ is a symmetric $m\times m$ matrix, and $B_S$ is its $2N\times2N$ submatrix with rows and columns in $S$. When all the modes are equally squeezed and $m$ is large enough compared to $N$, the matrix $\sqrt m\,B_S/\tanh r$ is close to a symmetric matrix with i.i.d.\ circular complex Gaussian entries.

We therefore consider in this subsection a $2N\times2N$ symmetric matrix $M$, $M_{ij}=M_{ji}$, whose entries $M_{ij}$ with $i<j$ are i.i.d.\ with measure $\frac{d^2M_{ij}}{\pi}e^{-|M_{ij}|^2}$; the diagonal does not enter the Hafnian
\begin{equation}
  \hf(M)\;=\;\sum_{T\in\mathcal T_{2N}}\ \prod_{\{i,j\}\in T}M_{ij},
\end{equation}
where $\mathcal T_{2N}$ is the set of the $(2N-1)!!$ perfect matchings of $[2N]$. Since distinct matchings give orthogonal monomials, $\langle|\hf M|^2\rangle=(2N-1)!!$, and we normalize
\begin{equation}
  \label{eq:moment-hf}
  \langle Y^k\rangle\;=\;\frac{\langle|\hf M|^{2k}\rangle}{(2N-1)!!^{\,k}},
\end{equation}
so that $Y$ is the probability of a collision-free pattern of $2N$ photons in units of its average. Here $N$ is the number of pairs of photons.

Using the representation \eqref{eq:hf-contour} to the power $2k$, we introduce $2k\times2N$ variables $u_i^a$ with measure $d\mu_{2k}(u)=\prod_{a=1}^{2k}\prod_{i=1}^{2N}du_i^a/\big(2\pi i\,(u_i^a)^2\big)$, and, as for the permanent, we use the replicas $a\in[k]$ for the $k$ copies of $\hf M$ and $b\in[2k]\setminus[k]$ for the $k$ copies of $\overline{\hf M}$. Averaging with $\langle e^{xM_{ij}+y\overline{M}_{ij}}\rangle=e^{xy}$,
\begin{equation}
  \begin{split}
  \langle|\hf M|^{2k}\rangle&=\cint d\mu_{2k}(u)\;\exp\Big(\sum_{i<j}X^+_{ij}X^-_{ij}\Big),\\
  X^+_{ij}&=\sum_{a=1}^{k}u_i^a\,u_j^a,\qquad X^-_{ij}=\sum_{b=k+1}^{2k}u_i^b\,u_j^b .
  \end{split}
\end{equation}
Writing $\sum_{i<j}X^+_{ij}X^-_{ij}=\frac12\sum_{a,b}\big[R_{ab}^2-\sum_i(u_i^au_i^b)^2\big]$, the second term is divisible by squares and can be dropped, as explained after~\eqref{eq:averaged}, so that
\begin{equation}
  \label{eq:overlap-hf}
  \begin{split}
  \langle|\hf M|^{2k}\rangle&=\cint d\mu_{2k}(u)\;\exp\Big(\frac12\sum_{a=1}^{k}\sum_{b=k+1}^{2k}R_{ab}^2\Big),\\
  R_{ab}&=\sum_{i=1}^{2N}u_i^a\,u_i^b .
  \end{split}
\end{equation}
Compared with~\eqref{eq:overlap-c}, the product $R_{ab}S_{ab}$ of the row and column overlaps is replaced by $R_{ab}^2$.

We decouple~\eqref{eq:overlap-hf} with a real Hubbard-Stratonovich field. Defining $\E_t(\cdot)=\int\prod_{a,b}\frac{dt_{ab}}{\sqrt{2\pi}}\,e^{-\sum_{a,b}t_{ab}^2/2}(\cdot)$, we have $\E_t\,e^{\sum_{a,b}t_{ab}R_{ab}}=e^{\frac12\sum_{a,b}R_{ab}^2}$, the sites decouple, and the integral over the variables of each site is~\eqref{eq:per-contour} with $N\to k$ and $M\to t$. Therefore
\begin{equation}
  \label{eq:hf-dual}
  \langle|\hf M|^{2k}\rangle\;=\;\E_t\big[(\per t)^{2N}\big],
\end{equation}
where $t$ is a $k\times k$ matrix of i.i.d.\ real standard Gaussians.

The right-hand side is the $2N$-th moment of the permanent of a real Gaussian $k\times k$ matrix, and~\eqref{eq:hf-dual} is the identity~\eqref{eq:real-dual} of the real case with $N$ and $k$ exchanged: in~\eqref{eq:real-dual}, $\E_z$ is precisely the average over a $2k\times2k$ symmetric matrix with i.i.d.\ circular complex Gaussian entries. The moments of the permanents of real Gaussian matrices and the moments of the Hafnians of symmetric complex Gaussian matrices are thus the same function of two variables, $\langle(\per M)^{2k}\rangle_N=\langle|\hf M|^{2N}\rangle_{2k}$, where the subscript denotes the size of $M$, real in the first case and symmetric complex in the second.
The Hafnian–permanent connection was also investigated in Refs.~\cite{PhysRevLett.134.140601,PhysRevA.111.042412,shou2026anticoncentrationentanglementgaussianboson}, where both moments and benchmarking were considered.

The real case of the previous section is the limit of large matrices at fixed moment of this function; Gaussian boson sampling requires the opposite limit, the large moments of the permanent of a real matrix of fixed size $k$. We checked~\eqref{eq:hf-dual} exactly for small sizes, for instance $\langle|\hf M|^{6}\rangle=\E_t(\per t)^8=40279680$ for $2N=8$ and $k=3$.

For $k=1$, \eqref{eq:hf-dual} gives $\E\,t^{2N}=(2N-1)!!$. For $k=2$, $\per t=t_{11}t_{22}+t_{12}t_{21}$ is the sum of two independent products of Gaussians, and one obtains the exact result:
\begin{equation}
  \label{eq:k2-hf}
  \langle Y^2\rangle\;=\;\sum_{l=0}^{N}\binom{N}{l}^2\binom{2N}{2l}^{-1}=\frac{4^N}{\binom{2N}{N}}\;\simeq\;\sqrt{\pi N}.
\end{equation}

For general $k$, rescaling $t_{ab}=\sqrt N\,s_{ab}$ as before,
\begin{equation}
  \begin{split}
  \langle|\hf M|^{2k}\rangle&=(2\pi)^{-k^2/2}N^{kN+k^2/2}\int\prod_{a,b}ds_{ab}\;e^{-N\Phi_{\hf}({\bm s})},\\
  \Phi_{\hf}&=\frac12\sum_{a,b}s_{ab}^2-2\ln|\per({\bm s})| .
  \end{split}
\end{equation}
The stationarity condition reads $s_{ab}=2\,\partial_{ab}\per({\bm s})/\per({\bm s})$, homogeneity gives $\sum_{a,b}s_{ab}^2=2k$, and the replica-symmetric saddle is $s_{ab}=s_0$ with $s_0^2=2/k$. Dividing by $(2N-1)!!^{\,k}=\big[\sqrt2\,(2N/e)^N\big(1+O(1/N)\big)\big]^k$, the {\it Hafnian} rate function is
\begin{equation}
  \label{eq:rate-hf}
  \lim_{N\to\infty}\frac1N\ln\langle Y^k\rangle\;=\;2\ln\Gamma(k+1)-k\ln k\;=\;\Lambda_{\mathbb C}(k),
\end{equation}
the same as for the {\it permanents} of complex Gaussian matrices, with $N$ now the number of pairs.

The prefactor is different. Since the field is real, the continuous phase orbit of the complex case reduces to the sign changes $s_{ab}\to\varepsilon_a\eta_b\,s_{ab}$, with $\varepsilon_a,\eta_b=\pm1$, which leave $\per({\bm s})^2$ invariant: there are $2^{2k-1}$ isolated saddle points there is no moduli space volume to compute. In the variables $s_{ab}=s_0(1+x_{ab})$, $\Phi_{\hf}$ coincides with $\Phi_{\mathbb C}$ of the previous subsection restricted to real $x_{ab}$, so that the Hessian is~\eqref{eq:hessian-complex} without the imaginary parts, and we find (see the Appendix)
\begin{equation}
  \label{eq:result-hf}
  \langle Y^k\rangle\;=\;C^{\hf}_k\,\big(1+a^{\hf}_1/N+O(1/N^2)\big)\,e^{N\Lambda_{\mathbb C}(k)},
\end{equation}
with
\begin{equation}
  \label{eq:Ck-hf}
  \begin{split}
  C^{\hf}_k&=2^{\frac{k-1}{2}}\Big(\frac{k-1}{k-2}\Big)^{(k-1)^2/2},\\
  a^{\hf}_1(k)&=\frac{k(k-1)(k^2-2k+2)}{12\,(k-2)^3} .
  \end{split}
\end{equation}
The absence of a power of $N$ reflects the absence of a moduli space. As for the permanents, $C^{\hf}_k$ and $a^{\hf}_1$ are singular at $k=2$, where the Hessian has an additional flat direction; here it is an arc, whose integration gives the $\sqrt{\pi N}$ of~\eqref{eq:k2-hf}.

In Table~\ref{tab:hf} we compare~\eqref{eq:result-hf} with exact moments for $k=3$. For a $3\times3$ real matrix, $\E_t(\per t)^{2N}$ can be computed exactly for large $N$: expanding $\per t$ along the first row, which is Gaussian given the other two, gives $(2N-1)!!\,\E|C|^{2N}$, where $C$ is the vector of the $2\times2$ permanents of the last two rows; $C$ is linear in the third row, and its Gaussian average gives $N!\,[\sigma^N]\big(1-4p\sigma+4p^2\sigma^2-8q^2\sigma^3\big)^{-1/2}$, with $p=\sum_jt_{2j}^2$ and $q=2t_{21}t_{22}t_{23}$, whose average over the second row is elementary. Including the factor $1+a^{\hf}_1/N$, the deviation drops below $0.2\%$ at $N=100$.

\begin{table*}
\caption{Moments $\langle Y^3\rangle=\langle|\hf M|^{6}\rangle/(2N-1)!!^{\,3}$ for the Hafnians of symmetric complex Gaussian matrices of size $2N$: exact values, leading saddle-point prediction $C^{\hf}_3e^{N\Lambda_{\mathbb C}(3)}=8\,(4/3)^N$ from~\eqref{eq:result-hf}, and the same including the factor $1+a^{\hf}_1/N$, with $a^{\hf}_1(3)=5/2$. Relative deviations from the exact values in parentheses.}
\label{tab:hf}
\begin{ruledtabular}
\begin{tabular}{clll}
$N$ & exact & leading & with $1/N$ \\
\hline
10  & $210.999$                & $142.062$ ($-32.7\%$)               & $177.577$ ($-15.8\%$) \\
20  & $3053.37$                & $2522.69$ ($-17.4\%$)               & $2838.03$ ($-7.1\%$) \\
40  & $8.55960\times10^{5}$    & $7.95499\times10^{5}$ ($-7.1\%$)    & $8.45217\times10^{5}$ ($-1.3\%$) \\
100 & $2.56104\times10^{13}$   & $2.49439\times10^{13}$ ($-2.6\%$)   & $2.55675\times10^{13}$ ($-0.2\%$) \\
300 & $2.44564\times10^{38}$   & $2.42500\times10^{38}$ ($-0.8\%$)   & $2.44520\times10^{38}$ ($-0.02\%$) \\
\end{tabular}
\end{ruledtabular}
\end{table*}

\section{From moments to distributions}

With all moments known analytically, one could invert the Mellin/Laplace transform to obtain the distribution $P(Y)$ of a permanent or a Hafnian, appropriately normalized. We could not compute all the moments; in particular, those with $k\leq 2$ are inaccessible because the RS result fails and we must break the RS at the saddle point. The analytic expressions for $\Lambda(k)$ we obtained apply only for $k>2$, since at $k=2$ the saddle-point method fails. This manifests itself also in the fact that $\Lambda(k)<0$ when $1<k<2$ while the data show $\lim_{N\to\infty}\ln\langle Y^k\rangle/N=0$ here. So we can only reliably invert the Mellin transform for large values of $Y$, where the main contribution to the integral falls in a region of $k>2$ where our results hold.

In this section, as a representative example for which it is easier to obtain the exact numerical values, we look at the permanents of the case $M_{ij}=\pm 1$. There is no reason to expect that this would not hold {\it mutatis mutandis} for all the other cases examined in this paper.

Quantitatively, one first writes the inversion Mellin formula:
\begin{equation}
    P(y)=\int_{\mathcal{B}} \frac{dk}{2\pi i} \langle Y^{k}\rangle y^{-k-1},
\end{equation}
where Bromwich path $\mathcal{B}$ passes to the right of all the singularities of the integrand. This step is the analogue of inverting a {\it canonical} distribution $e^{-\beta H}$ to get a {\it microcanonical} distribution $\delta(E-H)$ by doing a complex contour integral over $\beta$. In this analogy $k\leftrightarrow \beta$ and $-\ln y\leftrightarrow H$, the Hamiltonian, while $P(y)$ plays the role of the density of states.

For exponentially large $y$, one can define $\eta=\frac{1}{N}\ln y =O(1)$ (as an energy density should be) and use the asymptotic formula for the moments to obtain
\begin{equation}
    P(y)\simeq\frac{1}{y}\,\frac{1}{2\pi i}\int_{c-i\infty}^{c+i\infty}C_k\,N^{k/2}\,e^{N[\Lambda(k)-k\eta]}\,dk.
\end{equation}

The integral can be computed by deforming the contour to pass through the saddle point $k^*$, obtaining
\begin{equation}
P(y)\simeq\frac{1}{y}\,\frac{C_{k^*}\,N^{k^*/2}\,e^{N[\Lambda(k^*)-k^*\eta]}}{\sqrt{2\pi N\Lambda''(k^*)}},
\label{eq:Py_saddle}
\end{equation}
with the saddle point equation:
\begin{equation}
    \Lambda'(k^*)=\eta=\frac{1}{N}\ln y,
\end{equation}
as long as $k^*\geq 2$. This means we must have $\eta>\eta_c=\Lambda'(2)$. Numerically, $\eta_c=\Lambda'(2)=0.36066$ for $\mathbb{R}$ and $0.15242$ for the $\mathbb{C}$ ensembles.

Working at  leading exponential order, $P(y)\simeq e^{-N (I(\eta)+\eta)}$ where parametrically
\begin{equation}
\begin{split}
I(\eta)&=
\begin{cases}
2\eta, & 0<\eta\le\eta_c,\\[4pt]
2k\,\psi\!\left(k+\tfrac12\right)-2\ln\Gamma\!\left(k+\tfrac12\right) & \\
\quad+\ln\pi-\dfrac{2k^2}{2k-1}, & \eta>\eta_c,
\end{cases}\\[4pt]
\eta&=2\ln2+2\psi\!\left(k+\tfrac12\right)-\ln(2k-1)-\frac{2k}{2k-1}.
\end{split}
\end{equation}
We identify two regimes: $0<\eta<\eta_c$ (keeping in mind that this excludes $y=O(1)$ which would be $\eta=O(1/N)$) and $\eta>\eta_c$.  So for $\eta\leq \eta_c$, or $y<e^{N\eta_c}$ we have $NI(\eta)=2N\eta=2\ln y$ so
\begin{equation}
    P(y)\sim\frac{1}{y}e^{-NI(\eta)}\sim \frac{A_N}{y^3},
\end{equation}
Going back to the random variable $p=\per M$, this is a $P(p)\sim1/p^5$ power-law. Notice that we cannot compute $A_N$ nor other details of this distribution (like terms $\ln y$ etc) since we cannot simply analytically continue the moments $\langle Y^k\rangle$ to $k<2$. In fact, although we know that $\Lambda(k)=0$ here, the expressions of the $C_k$ do not make any sense for $k<2$, in particular they become complex. This power law does not exist in a separate regime but must be understood as a limiting form when one approaches the bulk from the tail. The precise form of the bulk distribution is subject to a proper RSB calculation, to be left as future work.

\begin{figure*}
    \includegraphics[width=0.9\columnwidth]{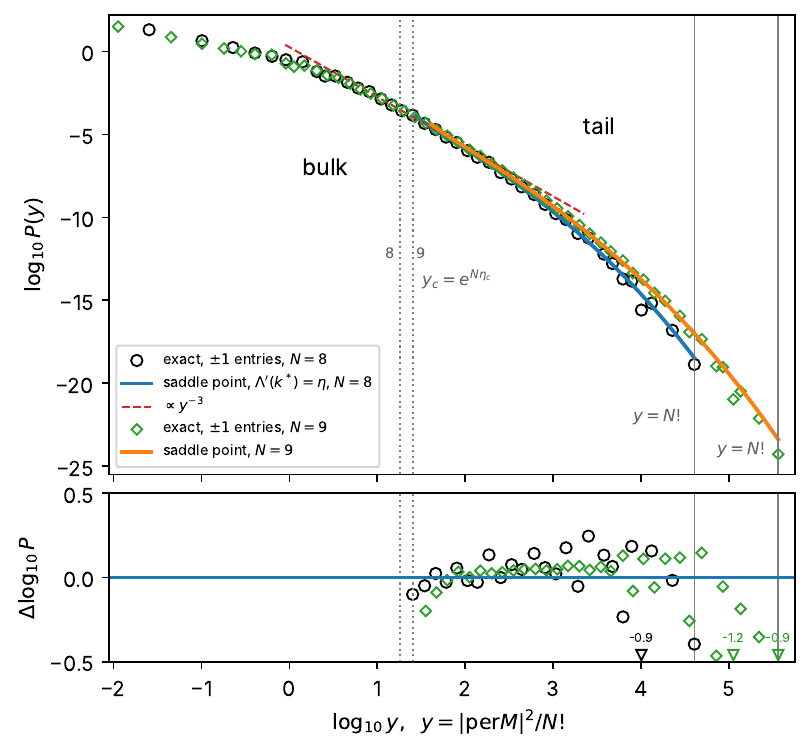}
    \includegraphics[width=0.9\columnwidth]{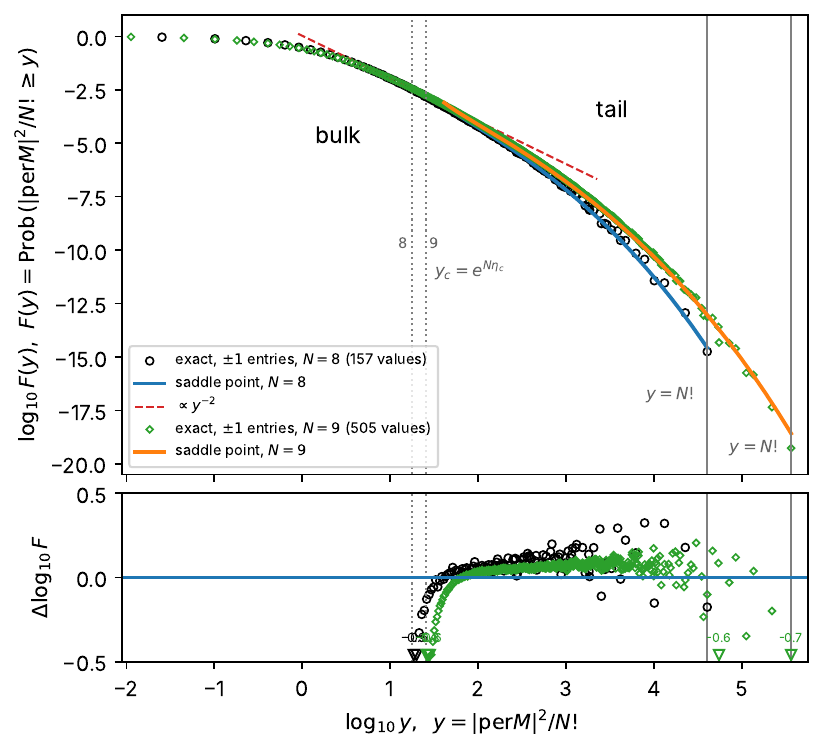}
    \caption{The full counting of the $N=8,9$, $M_{ij}=\pm 1$ normalized permanents of random matrices ($N=8$, circles; $N=9$, diamonds) versus the (parameter free) saddle point calculation of the probability density $P(y)$ (left) in \eqref{eq:Py_saddle} and of the cumulative $F(y)=\int_y^{+\infty} dy' P(y')$ (right). The bulk is not explained by the replica symmetric saddle point, while the incipient power law $P\sim y^{-3}, F\sim y^{-2}$ and the stretched exponential tails are. The differences between the numerical counting and the saddle-point formula are plotted in the lower panels.}
    \label{fig:PyN8pm1}
\end{figure*}

For $\eta>\eta_c$ and $\eta\ll \ln N$ (the location of the wall),
using the asymptotic form of $I(\eta)$ we find
\begin{equation}
    P(y)\sim e^{-eN y^{1/N}/2},
\end{equation}
a stretched exponential with a very small stretching exponent $1/N$. So, close  to --but not exactly-- a power law, which curvature in a $log-log$ plot is appreciable, in the appropriate scale. The saddle point result \eqref{eq:Py_saddle} is plotted against exact enumeration of the normalized permanents the $\pm 1$ random real matrices in Figure \ref{fig:PyN8pm1}. 

\section{Implications for the experiments}

\begin{figure}
    \includegraphics[width=0.95\columnwidth]{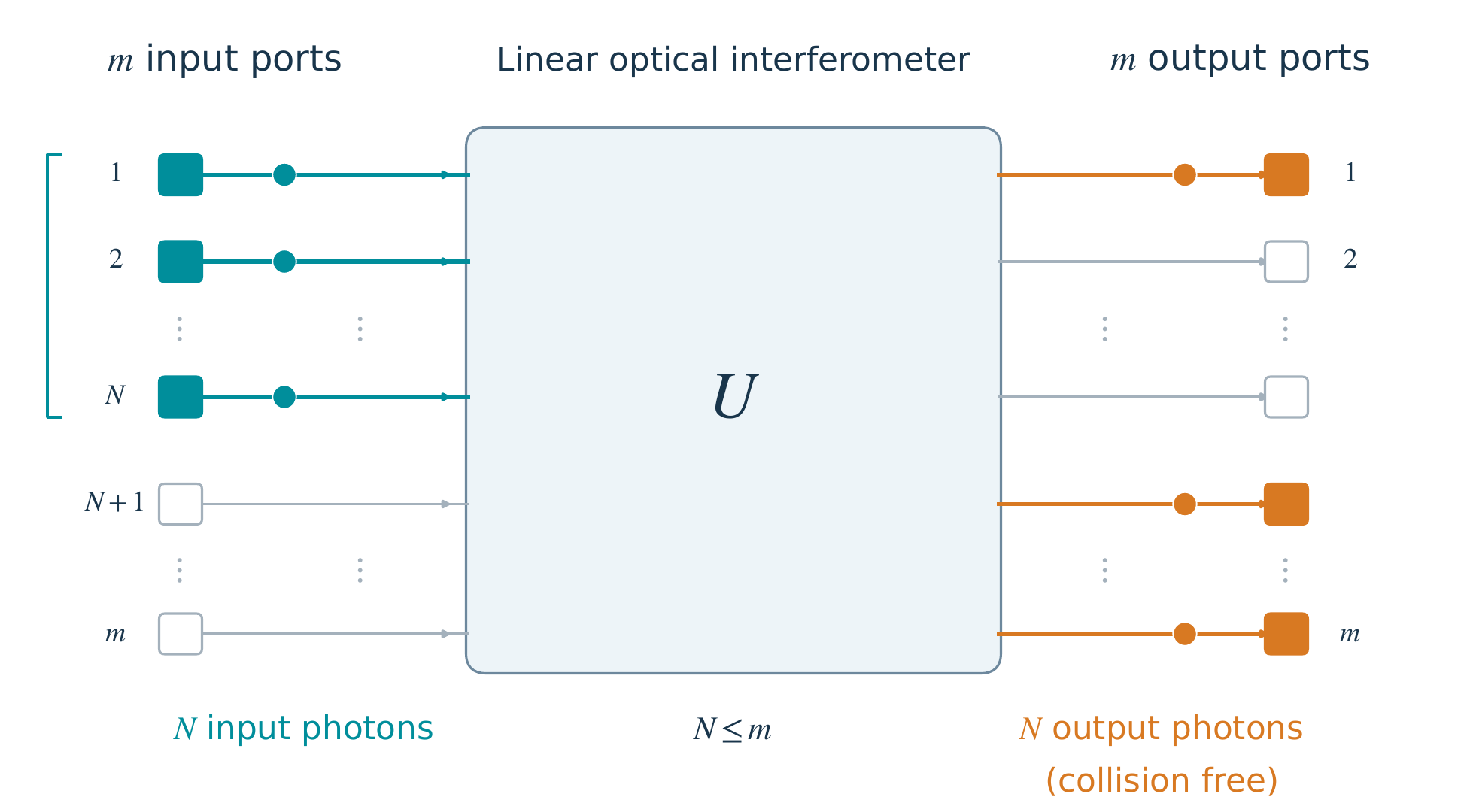}
    \caption{$N$ single photons, namely  photon-number states, illuminate $N$ of the $m$ input modes of a Haar-random interferometer $U$. The output distribution is expressed in terms of permanents.}
    \label{fig:boson_sampling_distributed_outputs}
\end{figure}

{\bf Boson sampling with single photons}--- In boson sampling~\cite{AaronsonArkhipov2011} $N$ single photons, i.e.\ photon-number states, are sent into $N$ of the $m$ input modes of a Haar-random interferometer $U$. See Fig.\ \ref{fig:boson_sampling_distributed_outputs}.

The probability of a collision-free outcome, one photon in each output mode of a set $S$, is $p_S=|\per U_S|^2$, where $U_S$ is the $N\times N$ submatrix of $U$ with rows in $S$ and columns the occupied input modes. For $m$ large enough compared to $N$, $M=\sqrt m\,U_S$ is close to the complex Gaussian matrices of the previous section, and
\begin{equation}
  Y_S\;=\;\frac{|\per M|^2}{N!}\;\simeq\;\mathcal C\,p_S,\qquad \mathcal C=\binom mN,
\end{equation}
is the probability of the outcome in units of that of the uniform distribution on the $\mathcal C$ collision-free outcomes. The collision-free outcomes carry most of the probability as long as $m\gg N^2$: their total weight, averaged over $U$, is $\binom mN\langle|\per U_S|^2\rangle=\prod_{j=0}^{N-1}\frac{m-j}{m+j}\simeq e^{-N^2/m}$. Averaging over $U$, or over the outcomes for a fixed interferometer, $\langle Y^k\rangle$ is the $k$-th moment of the output probabilities, and the results of the previous section are statements about the shape of the output distribution of the experiment.

We consider the linear cross-entropy score
$\mathcal{C}\sum_S q_S p_S$ of the observed distribution $q_S$,
which requires the ideal probabilities $p_S$ of the observed
outcomes. A general framework for computing ensemble-averaged
scores and second moments in photonic sampling experiments,
including beyond the dilute regime, was developed in
Ref.~\cite{kolarovszki2026generalframeworkanticoncentrationlinear}.
Averaged over $U$, it equals $\langle Y^2\rangle=N+1$ for an ideal boson sampler, $1$ for a uniform sampler, and $\sum_{j=0}^N1/j!\to e$ for fully distinguishable photons, whose output probabilities are $q_S=\per(|U_S|^2)$: in this last case the phase invariance of the entries of $M$ forces the two permutations in $|\per M|^2$ to coincide, and the average reduces to that of $2^{\#\text{fixed points}}$ over random permutations.

For comparison, in random circuit sampling the output probabilities follow the Porter-Thomas distribution, $\langle Y^k\rangle=k!$, with an ideal score equal to $2$ and a vanishing rate function: the output distribution of boson sampling is much more heavy-tailed.

The higher moments determine the statistical cost of such tests. The score is estimated by the average of $Y_S$ over $\mathcal M$ samples, with variance $\langle Y^3\rangle-\langle Y^2\rangle^2$ per sample, so that a relative precision $\varepsilon$ requires
\begin{equation}
  \label{eq:samples-bs}
  \begin{split}
  \mathcal M&\simeq\Big(\frac{\langle Y^3\rangle}{\langle Y^2\rangle^2}-1\Big)\varepsilon^{-2}\\
  &\simeq\Big[\frac{32\pi}{9\sqrt3}\,\frac{N\,(4/3)^N}{(N+1)^2}\Big(1+\frac2N\Big)-1\Big]\varepsilon^{-2},
  \end{split}
\end{equation}
where we used~\eqref{eq:k2-complex} and~\eqref{eq:result-complex} with $C^{\mathbb C}_3=32\pi/(9\sqrt3)$, $\Lambda_{\mathbb C}(3)=\ln\frac43$ and $a^{\mathbb C}_1(3)=2$. 

With the exact values of Table~\ref{tab:complex}, the factor in brackets is about $10^2$ at $N=20$, $1.6\times10^4$ at $N=40$ and $3.4\times10^6$ at $N=60$. More generally, estimating $\langle Y^k\rangle$ from samples requires a number of samples growing as $e^{N[\Lambda_{\mathbb C}(2k-1)-2\Lambda_{\mathbb C}(k)]}$. Because of the heavy tails, the empirical average is far from Gaussian at these values of $\mathcal M$, since the square of its skewness grows as $e^{N[2\Lambda_{\mathbb C}(4)-3\Lambda_{\mathbb C}(3)]}\simeq e^{0.76N}$, and confidence intervals should rather be based on Chebyshev's inequality or on median-of-means estimators. These estimates assume that, for a fixed interferometer, the averages over the outcomes coincide with the averages over $U$, which is plausible given the number $\mathcal C$ of outcomes, but not proven. Finally, through a Legendre transform, $\Lambda_{\mathbb C}(k)$ gives the large deviations of $\frac1N\ln Y$, i.e.\ the probability of observing outcomes exponentially more likely than the uniform ones, which is what heavy-output tests probe.

The hardness of approximate boson sampling relies on the anti-concentration of the permanent~\cite{AaronsonArkhipov2011}, a statement about the left tail, $\Pr[Y<1/\mathrm{poly}(N)]$, which the integer moments do not determine. Convexity forces the rate function to vanish on $[0,2]$, the Paley-Zygmund inequality with $\langle Y^2\rangle=N+1$ gives only an anti-concentration of order $1/N$, and the higher moments, which grow exponentially, make such bounds worse rather than better.

\begin{figure}
    \includegraphics[width=0.95\columnwidth]{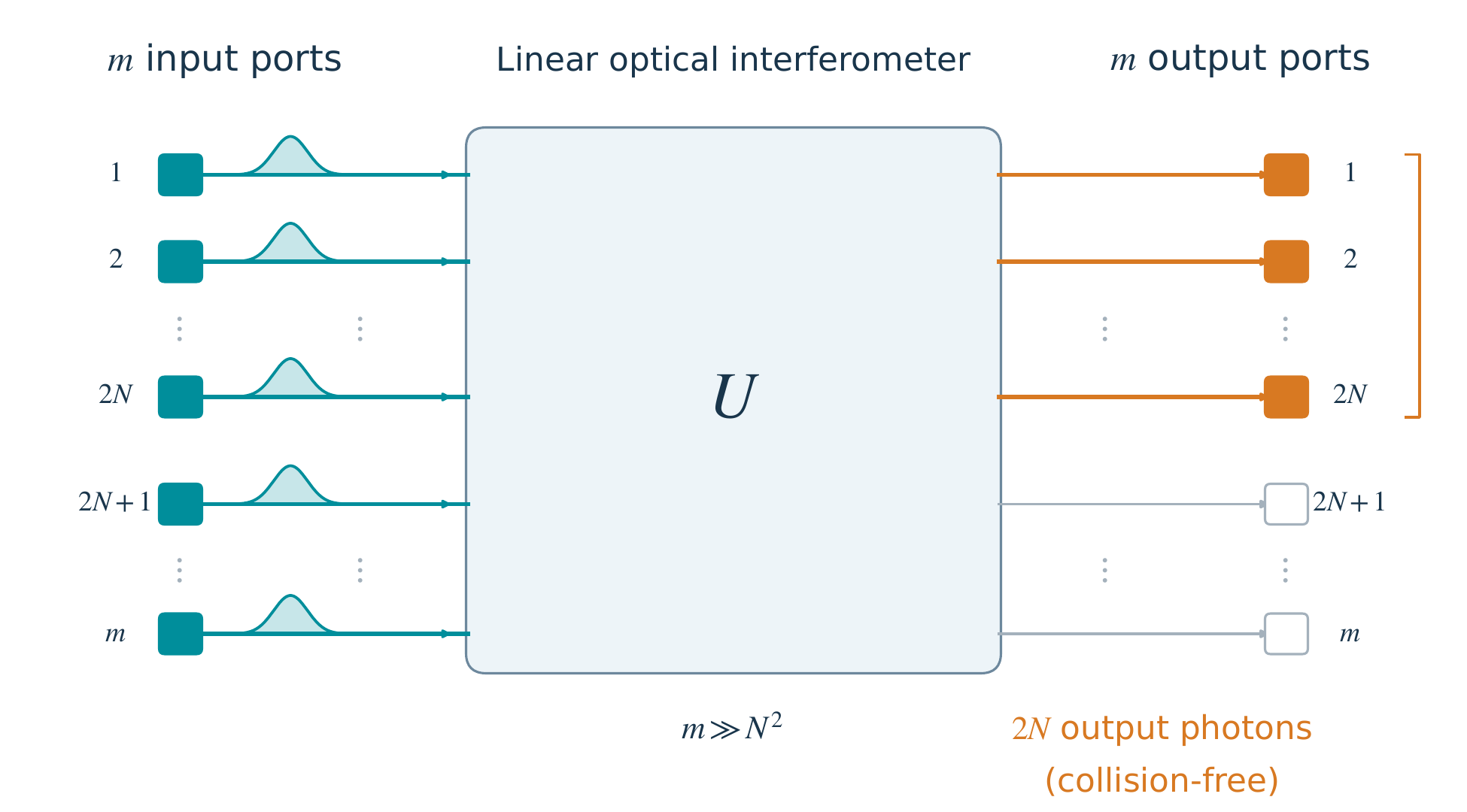}
    \caption{Identical single-mode squeezed vacuum states are injected into all $m$ input modes of a Haar-random interferometer $U$. We condition on detecting $2N$ photons in distinct output modes. The input photon number is not fixed and the output distribution is expressed in terms of Hafnians. }
    \label{fig:boson_sampling_individual_gaussians}
\end{figure}

{\bf Gaussian boson sampling}--- With squeezed states at the input, the number of photons is not fixed, and one conditions on detecting $2N$ photons in distinct modes. See Fig.\ \ref{fig:boson_sampling_individual_gaussians}.

The output probabilities are Hafnians, and we define $Y$ by~\eqref{eq:moment-hf}. The comparison with single photons is best done at equal number $n$ of detected photons, $n=N$ for boson sampling and $n=2N$ for Gaussian boson sampling:
\begin{itemize}
\item The ideal cross-entropy score is $\langle Y^2\rangle\simeq\sqrt{\pi n/2}$, Eq.~\eqref{eq:k2-hf}, instead of $n+1$.
\item By~\eqref{eq:rate-hf} the rate function is the same, $\Lambda_{\mathbb C}(k)$, but per pair of photons: at equal $n$ the moments grow as $e^{(n/2)\Lambda_{\mathbb C}(k)}$ instead of $e^{n\Lambda_{\mathbb C}(k)}$, the square root of the growth for single photons, and without a power of $n$ in front. The output distribution of Gaussian boson sampling is less heavy-tailed per detected photon.
\item Correspondingly, the number of samples needed to estimate the score is
\begin{equation}
  \label{eq:samples-gbs}
  \mathcal M\;\simeq\;\Big[\frac{8\,(4/3)^N}{\pi N}\Big(1+\frac{5}{2N}\Big)-1\Big]\varepsilon^{-2},
\end{equation}
which grows as $(2/\sqrt3)^n$ rather than $(4/3)^n$: with the exact moments, the factor in brackets is about $47$ for $n=40$ photons, against $1.6\times10^4$ for single photons.
\end{itemize}
In the calculation, the difference comes from the amplitude structure. For single photons, the amplitude involves both the input and the output modes, the rows and the columns of $U_S$; the replica calculation then involves both the row and the column overlaps, coupled as $R_{ab}S_{ab}$, which are decoupled by complex fields with a continuous phase symmetry, the origin of the moduli space and of the power $N^{(k-1)/2}$. 
Analytical results for linear cross-entropy
benchmarking of ideal Gaussian boson sampling were derived
in Ref.~\cite{PRXQuantum.5.040312}, and extensions to immanants in Ref.~\cite{Daigle_2025}.

For squeezed states, the photons are created in pairs, and the amplitude involves only the output modes; the overlaps appear as $R_{ab}^2$, which is decoupled by real fields with only a discrete sign symmetry. The results for Gaussian boson sampling assume that the output submatrices are close to symmetric complex Gaussian matrices, which requires many equally squeezed modes and $m\gg N^2$; with few squeezed sources, the matrix $B$ has low rank, and the ensemble is different. We do not consider losses, partial distinguishability, or threshold detectors here, and the anti-concentration of the Hafnian, needed for the hardness of Gaussian boson sampling, is not determined by the integer moments, as for the permanent.

\section{Conclusions}
\label{sec:conclusions}

We showed that an integral representation of the permanent can compute the normalized $k$-th moments $\langle Y^k\rangle$ of the permanent of random matrices, both real and complex. The calculation can be done by introducing Hubbard-Stratonovich fields, and continued with a saddle point analysis valid for large $N$. We found that the relevant saddle point in the HS fields is the replica symmetric one for all moments $k>2$ while for $k\leq 2$ the RS needs to be broken. We will do this in a future paper.

Using this technology, we could compute the overall prefactors and $1/N$ corrections to obtain, for Gaussian distributions, to $O(1/N)$:
\begin{equation}
    \langle |\per M|^{2k}\rangle=N!^k\,e^{N\Lambda_{\mathbb F}(k)}\,N^{\alpha_{\mathbb F}(k)}\,C^{\mathbb F}_k\left(1+\frac{a^{\mathbb F}_1(k)}{N}\right),
    \label{eq:nosimple}
\end{equation}
with $\mathbb F=\mathbb R$ for real and $\mathbb F=\mathbb C$ for complex matrices, $\alpha_{\mathbb R}(k)=k/2$ and $\alpha_{\mathbb C}(k)=(k-1)/2$, and get explicit expressions of $C^{\mathbb F}_k$, $a^{\mathbb F}_1$ and $\Lambda_{\mathbb F}$. The rate $\Lambda_{\mathbb F}$ and the power $\alpha_{\mathbb F}$ are universal, and depend only on the elements $M_{ij}$ being real or complex, while, for real entries, the fourth cumulant multiplies the prefactor by $e^{W_k}$, Eq.~\eqref{eq:Wk}, and $\kappa_4^2$ and $\kappa_6$ only modify the coefficient of $1/N$.

For the Hafnians of symmetric complex Gaussian matrices, relevant for Gaussian boson sampling, the same asymptotic expansion holds 
\begin{equation}
    \langle|\hf M|^{2k}\rangle=(2N-1)!!^{\,k}e^{N\Lambda_{\mathbb C}(k)}\, C^{\hf}_k\,\big(1+a^{\hf}_1/N\big),
    \label{eq:nonosimple}
\end{equation}
with the rate $\Lambda_{\mathbb C}$, no power of $N$ (as the saddles are isolated and no moduli space exists), and the constants $C^{\hf}_k$ and $a^{\hf}_1$. The moments of the permanents of complex Gaussian matrices are invariant under the exchange of the size of the matrix and of the order of the moment, while those of the permanents of real Gaussian matrices and of the Hafnians of symmetric complex Gaussian matrices are exchanged into each other.

\begin{acknowledgments}
We acknowledge support from INFN through the project ``QUANTUM" and from the Italian funding within the ``Budget MUR - Dipartimenti di Eccellenza 2023–2027" - Quantum Sensing and Modelling for One-Health (QuaSiModO).
We acknowledge support from Projects PN-RIC Q-SUD (Q-SUD B99H26000380007) and PN-RIC QUANTAS (SYNERGIA B99H26000410007).
PF acknowledges support from the Italian National Group of Mathematical Physics (GNFM-INdAM). 
\end{acknowledgments}

\appendix
\section*{Appendix}

{\bf Compact representations}---
The formula \begin{equation}
    \det (M)=\int \prod_{i=1}^N d\psi_i d\overline{\psi}_i \exp\Big(\sum_{i,j=1}^N \overline{\psi}_i\,M_{ij}\,\psi_j\Big)
\end{equation}
has a direct generalization to permanents:
\begin{equation}
    \operatorname{per} (M)=\int \prod_{i=1}^N d\phi_i d\overline{\phi}_i \exp\Big(\sum_{i,j=1}^N \overline{\phi}_i\,M_{ij}\,\phi_j\Big) \,,
\end{equation}
where the $\phi_i$ are nilpotent commuting variables, called zeons. We used the representation in terms of contour integrals, because it is more intuitive and familiar.

We notice en passant that one has
\begin{equation}
    \operatorname{pf}  (M)=\int \prod_{i=1}^N d\psi_i  \exp\Big(\sum_{i,j=1}^N {\psi}_i\,M_{ij}\,\psi_j\Big)\,,
\end{equation}
and 
\begin{equation}
    \operatorname{hf}  (M)=\int \prod_{i=1}^N d\phi_i  \exp\Big(\sum_{i,j=1}^N {\phi}_i\,M_{ij}\,\phi_j\Big)\,.
\end{equation}
So determinants, permanents, Pfaffians, and Hafnians have similar compact representations if we introduce nilpotent variables.

{\bf Saddle-point expansion}--- The Gaussian moments of the real and of the complex case have the common form
\begin{equation}
  \label{eq:app-integral}
  \begin{split}
  &N^{kN+D}\int\prod_{e}\frac{d^2y_e}{\pi}\;e^{-N\Phi({\bm y})},\\
  &\Phi({\bm y})=\sum_e|y_e|^2-\ln\Omega({\bm y})-\ln\overline{\Omega({\bm y})}.
  \end{split}
\end{equation}
The fields $y_e$ live on the edges $e$ of a replica graph $\Gamma$. For real entries $\Gamma$ is the complete graph on $[2k]$, $e=\{a,b\}$, $D=k(2k-1)$ and $\Omega=\hf$; for complex entries $\Gamma=K_{k,k}$, $e=(a,b)$ with $a\in[k]$ and $b\in[2k]\setminus[k]$, $D=k^2$ and $\Omega=\per$. In both cases $\Omega({\bm y})=\sum_T\prod_{e\in T}y_e$ is the sum over the perfect matchings $T$ of $\Gamma$, and the replica-symmetric saddle point is $y_e=y_0$, with $y_0^2=q_0$ the inverse of the degree of $\Gamma$, Eqs.~\eqref{eq:field-y0} and~\eqref{eq:y0-complex}; in this Appendix we write $q_0$ in both cases. We set $y_e=y_0(1+x_e)$ and use
\begin{equation}
  \frac{\Omega\big(y_0(\bm 1+x)\big)}{\Omega(y_0\bm 1)}\;=\;\sum_{n\ge0}\omega_n\,m_n(x),
\end{equation}
where $m_n(x)$ is the sum of $\prod_{e\in S}x_e$ over the sets $S$ of $n$ pairwise vertex-disjoint edges of $\Gamma$, and $\omega_n$ is the fraction of perfect matchings of $\Gamma$ that contain a given such set: $\omega_n=(2k-2n-1)!!/(2k-1)!!$ for real entries and $\omega_n=(k-n)!/k!$ for complex entries. Since $\omega_1=q_0$, the linear terms cancel and
\begin{equation}
  \label{eq:app-Phi}
  \begin{split}
  \Phi({\bm y})-\Phi(\bm y_s)&=q_0\sum_e|x_e|^2-2\Re\Big[\varphi(x)-q_0\sum_ex_e\Big],\\
  \varphi(x)&=\ln\sum_{n\ge0}\omega_n\,m_n(x)\\
  &=\sum_{n\ge1}\frac{1}{n!}\sum_{e_1,\dots,e_n}f^{(n)}_{e_1\dots e_n}\,x_{e_1}\cdots x_{e_n}.
  \end{split}
\end{equation}
The coefficients $f^{(n)}$ are the joint cumulants of $p^{(n)}_{e_1\dots e_n}$, equal to $\omega_n$ if the edges $e_1,\dots,e_n$ are pairwise vertex-disjoint and to $0$ otherwise:
\begin{equation}
  \label{eq:app-cumulants}
  \begin{split}
  f^{(2)}_{ef}&=p^{(2)}_{ef}-q_0^2,\\
  f^{(3)}_{efg}&=p^{(3)}_{efg}-q_0\big(p^{(2)}_{ef}+p^{(2)}_{eg}+p^{(2)}_{fg}\big)+2q_0^3,\\
  f^{(4)}_{efgh}&=p^{(4)}_{efgh}-q_0{\textstyle\sum_{[4]}}p^{(3)}-{\textstyle\sum_{[3]}}p^{(2)}p^{(2)}\\
  &\quad+2q_0^2{\textstyle\sum_{[6]}}p^{(2)}-6q_0^4,
  \end{split}
\end{equation}
where $\sum_{[n]}$ is the sum over the $n$ distinct ways of distributing the indices $e,f,g,h$.

{\bf Hessian and moduli space}--- Let $\mathcal P_0$ be the orthogonal projector, in the space of functions on the edges, on the functions of the form $x_e=\alpha+\sum_{v\in e}\xi_v$, with $v$ the vertices of $\Gamma$, and $\mathcal P_1=1-\mathcal P_0$. Explicitly,
\begin{equation}
  \begin{split}
  (\mathcal P_0)_{ef}&=\frac{|e\cap f|-2q_0}{2k-2}\quad\text{(real)},\\
  (\mathcal P_0)_{(ab),(a'b')}&=\frac1{k^2}+\frac{\delta_{aa'}-\frac1k}{k}+\frac{\delta_{bb'}-\frac1k}{k}\quad\text{(complex)},
  \end{split}
\end{equation}
of rank $n_0=2k$ and $n_0=2k-1$ respectively (for $K_{k,k}$, shifting $\xi_v$ by $+c$ on one group of vertices and by $-c$ on the other leaves $x$ unchanged). The quadratic part of~\eqref{eq:app-Phi} is diagonal in this decomposition. For real entries, writing $x_e=\alpha+\xi_a+\xi_b+\zeta_e$ for $e=\{a,b\}$, with $\sum_{a=1}^{2k}\xi_a=0$ and $\mathcal P_0\zeta=0$,
\begin{equation}
  \label{eq:hessian-real}
  \begin{split}
  &\Phi({\bm y})-\Phi(\bm{y}_s)\;=\;2k\,(\Re\alpha)^2+\frac{4(k-1)}{2k-1}\sum_{a=1}^{2k}(\Re\xi_a)^2\\
  &\quad+\frac{2(k-2)}{(2k-1)(2k-3)}\sum_e(\Re\zeta_e)^2\\
  &\quad+\frac{2(k-1)}{(2k-1)(2k-3)}\sum_e(\Im\zeta_e)^2+O(x^3).
  \end{split}
\end{equation}
For complex entries, writing $x_{ab}=\alpha+\xi_a+\eta_b+\zeta_{ab}$, with $\sum_{a=1}^k\xi_a=\sum_{b=k+1}^{2k}\eta_b=0$ and $\sum_{a=1}^k\zeta_{ab}=\sum_{b=k+1}^{2k}\zeta_{ab}=0$,
\begin{equation}
  \label{eq:hessian-complex}
  \begin{split}
  \Phi_{\mathbb C}({\bm y})-\Phi_{\mathbb C}(\bm{y}_s)\;=\;&2k\,(\Re\alpha)^2+2\sum_{a=1}^k(\Re\xi_a)^2\\
  &+2\sum_{b=k+1}^{2k}(\Re\eta_b)^2\\
  &+\frac{k-2}{k(k-1)}\sum_{a,b}(\Re \zeta_{ab})^2\\
  &+\frac{1}{k-1}\sum_{a,b}(\Im \zeta_{ab})^2+O(x^3).
  \end{split}
\end{equation}
In both cases the real part of $\mathcal P_0x$ enters with coefficient $2q_0$, i.e.\ as $2q_0\sum_e\big((\mathcal P_0\Re x)_e\big)^2$, and the real and imaginary parts of $\mathcal P_1x$ with coefficients $\lambda_P$ and $\lambda_Q$, read off from~\eqref{eq:hessian-real} and~\eqref{eq:hessian-complex}. Both $\lambda_P$ vanish at $k=2$, where the prefactors are singular.

The imaginary part of $\mathcal P_0x$ is flat. These $n_0$ directions generate the moduli space of the saddle, the orbit $y_e\to e^{i\sum_{v\in e}\theta_v}y_e$, which multiplies $\Omega({\bm y})$ by $e^{i\sum_v\theta_v}$, since every perfect matching covers each vertex once, and therefore leaves $|\Omega({\bm y})|$ and $\sum_e|y_e|^2$ invariant. For real entries, the moduli space is the torus $\theta\in[0,2\pi)^{2k}$ modulo the shift of all the $\theta_a$ by $\pi$, which acts trivially; for complex entries the common shift $\theta_a\to\theta_a+\theta$, $\theta_b\to\theta_b-\theta$ acts trivially, and the torus has dimension $2k-1$. Their volumes in the metric $\sum_e|dy_e|^2$ are
\begin{equation}
  \begin{split}
  V_0&=2^{k-1}(2\pi)^{2k}\Big(\frac{k-1}{2k-1}\Big)^{k-\frac12}\quad\text{(real)},\\
  V_0^{\mathbb C}&=(2\pi)^{2k-1}k^{-\frac12}\quad\text{(complex)}.
  \end{split}
\end{equation}
In the coordinates $y_e$, the Hessian of $\Phi$ at the saddle has, besides the $n_0$ zero modes, the eigenvalue $4$ with multiplicity $n_0$, and the eigenvalues $2\lambda_P/q_0$ and $2\lambda_Q/q_0$, each with multiplicity $D-n_0$. The product of its non-zero eigenvalues is
\begin{equation}
\begin{split}
     {\det}'H&=2^{8k(k-1)}\biggl(\frac{(k-1)(k-2)}{(2k-3)^2}\biggr)^{k(2k-3)}\quad\!\!\text{(real)},\qquad \\
  {\det}'H_{\mathbb C}&=2^{2k^2}\biggl(\frac{k(k-2)}{(k-1)^2}\biggr)^{(k-1)^2}\quad\text{(complex)}.
\end{split}
\end{equation}
Integrating over the $r=2D-n_0$ Gaussian directions and over the moduli space, and dividing by $N!^{\,k}$ with Stirling's formula,
\begin{equation}
  \langle Y^k\rangle\;\simeq\;(2\pi)^{\frac{r-k}{2}}\pi^{-D}\,\frac{V_0}{\sqrt{{\det}'H}}\;N^{\frac{n_0-k}{2}}\,e^{N[k-\Phi(\bm y_s)]},
\end{equation}
which gives~\eqref{eq:field-result} with~\eqref{eq:Ck}, and~\eqref{eq:result-complex} with~\eqref{eq:Ck-complex}. The power $N^{(n_0-k)/2}$, i.e.\ $N^{k/2}$ and $N^{(k-1)/2}$, reflects the dimension of the moduli space, while the last factors of $C^{\mathbb R}_k$ and $C^{\mathbb C}_k$ come from the $D-n_0$ modes $\zeta$.

{\bf The $1/N$ correction}--- To go to the next order we parametrize the integration variables as $y_e=y_0\,e^{i\sum_{v\in e}\theta_v}(1+x_e)$, with $x$ restricted to the slice $\Im\mathcal P_0x=0$ orthogonal to the moduli space. The tangent vectors to the orbit at $y_0(\bm 1+x)$ are $i(1+x_e)\sum_{v\in e}\delta\theta_v$, and the Jacobian of the change of variables, relative to its value at $x=0$, is
\begin{equation}
  \frac{\mathcal J(x)}{\mathcal J(0)}\;=\;\det\big[\mathcal P_0\,(1+\Re x)\,\mathcal P_0\big],
\end{equation}
where $\Re x$ acts as a diagonal matrix and the determinant is taken on the image of $\mathcal P_0$. Rescaling $x\to x/\sqrt N$,
\begin{equation}
  \begin{split}
  &e^{-N[\Phi({\bm y})-\Phi(\bm y_s)]}\,\frac{\mathcal J(x)}{\mathcal J(0)}
  =e^{-\Phi_2(x)}\Big[1+\frac{L-\Phi_3}{\sqrt N}\\
  &\qquad+\frac{J_2-L\,\Phi_3-\Phi_4+\frac12\Phi_3^2}{N}+O(N^{-3/2})\Big],
  \end{split}
\end{equation}
where $\Phi_2$ is the quadratic part~\eqref{eq:hessian-real} or~\eqref{eq:hessian-complex}, $\Phi_n=-2\Re\varphi_n$ for $n\ge3$, with $\varphi_n$ the term of degree $n$ in~\eqref{eq:app-Phi}, and, expanding the determinant,
\begin{equation}
  \begin{split}
  L&=\sum_e(\mathcal P_0)_{ee}\Re x_e=n_0\Re\alpha,\\
  J_2&=\frac12L^2-\frac12\sum_{e,f}(\mathcal P_0)_{ef}^2\,\Re x_e\,\Re x_f .
  \end{split}
\end{equation}
Denoting by $\langle\cdot\rangle_0$ the Gaussian average with weight $e^{-\Phi_2}$ on the slice, the coefficient of $1/N$ in $\langle Y^k\rangle$, $a^{\mathbb F}_1$ with $\mathbb F=\mathbb R,\mathbb C$, is
\begin{equation}
  \label{eq:app-a1}
  a^{\mathbb F}_1\;=\;\langle J_2\rangle_0-\langle L\,\Phi_3\rangle_0-\langle\Phi_4\rangle_0+\tfrac12\langle\Phi_3^2\rangle_0-\frac{k}{12},
\end{equation}
where $-k/12$ comes from Stirling's formula for $N!^{\,k}$. With $\sigma_{P,Q}=1/(2\lambda_{P,Q})$, the propagators on the slice are
\begin{equation}
  \begin{split}
  \mathcal G_{ef}&\equiv\langle x_ex_f\rangle_0=\frac{1}{4q_0}(\mathcal P_0)_{ef}+(\sigma_P-\sigma_Q)(\mathcal P_1)_{ef},\\
  \mathcal K_{ef}&\equiv\langle x_e\bar x_f\rangle_0=\frac{1}{4q_0}(\mathcal P_0)_{ef}+(\sigma_P+\sigma_Q)(\mathcal P_1)_{ef}.\\
  \,
  \end{split}
\end{equation}

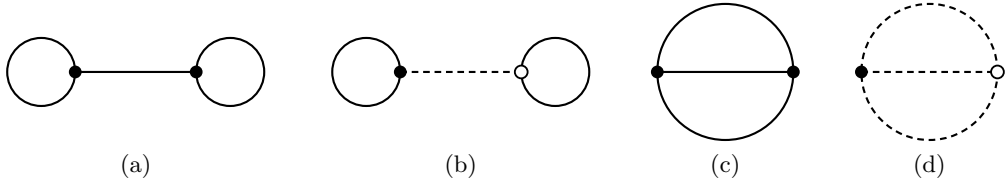
\begin{figure*}
\centering
\begin{tikzpicture}[
  hol/.style={circle,fill=black,inner sep=1.7pt},
  ahol/.style={circle,draw=black,fill=white,inner sep=1.7pt,line width=0.7pt},
  G/.style={line width=0.8pt},
  K/.style={line width=0.8pt,dash pattern=on 2.5pt off 1.8pt},
  lab/.style={font=\small}]
  % (a) dumbbell: two vertices of phi, bridge G
  \begin{scope}[shift={(0,0)}]
    \draw[G] (-0.45,0) circle (0.45);
    \draw[G] (2.05,0) circle (0.45);
    \draw[G] (0,0) -- (1.6,0);
    \node[hol] at (0,0) {}; \node[hol] at (1.6,0) {};
    \node[lab] at (0.8,-1.25) {(a)};
  \end{scope}
  % (b) dumbbell: vertices of phi and bar phi, bridge K
  \begin{scope}[shift={(4.3,0)}]
    \draw[G] (-0.45,0) circle (0.45);
    \draw[G] (2.05,0) circle (0.45);
    \draw[K] (0,0) -- (1.6,0);
    \node[hol] at (0,0) {}; \node[ahol] at (1.6,0) {};
    \node[lab] at (0.8,-1.25) {(b)};
  \end{scope}
  % (c) theta: two vertices of phi, three G lines
  \begin{scope}[shift={(7.7,0)}]
    \draw[G] (0.9,0) circle (0.9);
    \draw[G] (0,0) -- (1.8,0);
    \node[hol] at (0,0) {}; \node[hol] at (1.8,0) {};
    \node[lab] at (0.9,-1.25) {(c)};
  \end{scope}
  % (d) theta: vertices of phi and bar phi, three K lines
  \begin{scope}[shift={(10.4,0)}]
    \draw[K] (0.9,0) circle (0.9);
    \draw[K] (0,0) -- (1.8,0);
    \node[hol] at (0,0) {}; \node[ahol] at (1.8,0) {};
    \node[lab] at (0.9,-1.25) {(d)};
  \end{scope}
\end{tikzpicture}
\caption{Diagrams of order $1/N$ with two cubic vertices, contributing to $\frac12\langle\Phi_3^2\rangle_0$. Filled and open dots are the vertices $f^{(3)}$ of $\varphi$ and $\overline\varphi$; solid lines are the propagators $\mathcal G_{ef}=\langle x_ex_f\rangle_0$, equal to $\langle\bar x_e\bar x_f\rangle_0$ since $\mathcal G$ is real, and dashed lines are $\mathcal K_{ef}=\langle x_e\bar x_f\rangle_0$. (a,b) Dumbbell, $\frac14\sum_{e,f}v_e\mathcal G_{ef}v_f$ and $\frac14\sum_{e,f}v_e\mathcal K_{ef}v_f$, each loop being the tadpole $v_e$. (c,d) Theta, $\frac16\sum f^{(3)}_{efg}f^{(3)}_{e'f'g'}\mathcal G_{ee'}\mathcal G_{ff'}\mathcal G_{gg'}$ and $\frac16\sum f^{(3)}_{efg}f^{(3)}_{e'f'g'}\mathcal K_{ee'}\mathcal K_{ff'}\mathcal K_{gg'}$.}
\label{fig:diagrams}
\end{figure*}

\begin{widetext}
Since $\varphi$ is holomorphic, Wick's theorem involves only these two, and
\begin{equation}
  \begin{split}
  \langle J_2\rangle_0&=\frac14\sum_{e,f}\big[(\mathcal P_0)_{ee}(\mathcal P_0)_{ff}-(\mathcal P_0)_{ef}^2\big](\mathcal G+\mathcal K)_{ef},\\
  -\langle L\,\Phi_3\rangle_0&=\frac12\sum_{e,f}(\mathcal P_0)_{ee}(\mathcal G+\mathcal K)_{ef}\,v_f,\\
  -\langle\Phi_4\rangle_0&=\frac14\sum_{e,f,g,h}f^{(4)}_{efgh}\,\mathcal G_{ef}\mathcal G_{gh},\\
  \tfrac12\langle\Phi_3^2\rangle_0&=\frac14\sum_{e,f}v_e(\mathcal G+\mathcal K)_{ef}v_f\\
  &\quad+\frac16\sum f^{(3)}_{efg}f^{(3)}_{e'f'g'}\big(\mathcal G_{ee'}\mathcal G_{ff'}\mathcal G_{gg'}\\
  &\qquad\qquad+\mathcal K_{ee'}\mathcal K_{ff'}\mathcal K_{gg'}\big),
  \end{split}
\end{equation}
with $v_e=\sum_{f,g}f^{(3)}_{efg}\mathcal G_{fg}$. The two terms of $\frac12\langle\Phi_3^2\rangle_0$ are the dumbbell and the theta diagrams of Fig.~\ref{fig:diagrams}. Since $p^{(n)}$, $\mathcal G$ and $\mathcal K$ depend only on the coincidences among the replica indices of the edges involved, all the sums reduce to sums over coincidence patterns, which we evaluated exactly as rational functions of $k$. The results are given in Eq.~\eqref{eq:contrib-real} for real entries and in Eq.~\eqref{eq:contrib-complex} for complex entries,
\begin{equation}
  \label{eq:contrib-real}
  \begin{split}
  \langle J_2\rangle_0&=-\frac{k(2k-1)(2k-3)}{16(k-1)(k-2)},\qquad
  -\langle L\,\Phi_3\rangle_0=\frac{k\,(2k^2-8k+7)}{4(k-1)(k-2)},\\
  -\langle\Phi_4\rangle_0&=-\frac{k\,(10k^5-87k^4+296k^3-491k^2+399k-128)}{32(k-1)^3(k-2)^2},\\
  \text{dumbbell}&=\frac{k\,(2k^2-8k+7)^2}{32(k-1)^2(k-2)^2},\qquad
  \text{theta}=\frac{k\,(20k^6-68k^5-71k^4+564k^3-781k^2+314k+31)}{96(k-1)^3(k-2)^2(2k-5)},
  \end{split}
\end{equation}
\begin{equation}
  \label{eq:contrib-complex}
  \begin{split}
  \langle J_2\rangle_0&=-\frac{(k-1)^2}{4(k-2)},\qquad
  -\langle L\,\Phi_3\rangle_0=\frac{(2k-1)(2k^3-9k^2+10k-4)}{8k^2(k-2)},\\
  -\langle\Phi_4\rangle_0&=-\frac{(5k^2-6k+4)(2k^4-14k^3+35k^2-34k+12)}{32k^3(k-2)^2},\\
  \text{dumbbell}&=\frac{(2k^3-9k^2+10k-4)^2}{32k^3(k-2)^2},\qquad
  \text{theta}=\frac{5k^7+9k^6-113k^5+206k^4+12k^3-360k^2+336k-96}{48k^3(k-2)^3},
  \end{split}
\end{equation}
\end{widetext}
and their sums with $-k/12$ are~\eqref{eq:a1} and~\eqref{eq:a1-complex}. The factor $2k-5$ comes from $\omega_3=1/\big((2k-1)(2k-3)(2k-5)\big)$ in the theta diagram.

For large $k$ both $a^{\mathbb R}_1$ and $a^{\mathbb C}_1$ grow as $k/12$. In the complex case this growth is due entirely to the normalization by $N!^{\,k}$: the expansion of the normalization $\mathcal N_{N,k}$ used in~\eqref{eq:nezami-limit} contains the factor $1+(k^2-1)/(12kN)$, so that
\begin{equation}
  \begin{split}
  \frac{\langle|\per M|^{2k}\rangle}{\mathcal N_{N,k}}=\;&\Big(\frac{(k-1)^2}{k(k-2)}\Big)^{(k-1)^2/2}\\
  &\times\Big[1+\frac{2(k-1)^3}{3k(k-2)^3}\,\frac1N+O(N^{-2})\Big],
  \end{split}
\end{equation}
whose $1/N$ coefficient tends to $2/3$ as $k\to\infty$. We checked the propagators against a direct inversion of the Hessian, the Jacobian against a direct evaluation of the determinant, and the Wick contractions against a brute-force evaluation of the tensor sums for small $k$. For complex entries, $a^{\mathbb C}_1(3)=2$ agrees with the exact moments of Table~\ref{tab:complex}, extended to $N=100$ with the values of~\cite{Nezami2021}.

{\bf Hafnians}--- For the Hafnians of symmetric complex Gaussian matrices the dual integral~\eqref{eq:hf-dual} has real fields, and with $s_{ab}=s_0(1+x_{ab})$, $s_0^2=2/k$,
\begin{equation}
  \begin{split}
  &\Phi_{\hf}({\bm s})-\Phi_{\hf}(\bm s_s)\\
  &\quad=q_0\sum_{a,b}x_{ab}^2-2\Big[\varphi(x)-q_0\sum_{a,b}x_{ab}\Big],\qquad q_0=\frac1k,
  \end{split}
\end{equation}
with the same $\varphi$ as for complex permanents, now evaluated at real $x$. The quadratic part is~\eqref{eq:hessian-complex} without the imaginary parts: there are no zero modes, and the propagator is $\mathcal R_{ef}\equiv\langle x_ex_f\rangle_0=\frac1{4q_0}(\mathcal P_0)_{ef}+\sigma_P(\mathcal P_1)_{ef}=\frac12(\mathcal G+\mathcal K)_{ef}$. In the coordinates $s_{ab}$ the Hessian has the eigenvalue $2$ with multiplicity $2k-1$ and the eigenvalue $(k-2)/(k-1)$ with multiplicity $(k-1)^2$; summing over the $2^{2k-1}$ saddle points related by sign changes and dividing by $(2N-1)!!^{\,k}$ gives $C^{\hf}_k$ in~\eqref{eq:Ck-hf}. Without a moduli space there are no Jacobian terms, and the $1/N$ correction comes from the quartic vertex and from the dumbbell and theta diagrams only, which for real fields are those of Fig.~\ref{fig:diagrams}(a) and~(c):
\begin{equation}
  \begin{split}
  a^{\hf}_1\;=\;&\frac14\sum_{e,f,g,h}f^{(4)}_{efgh}\mathcal R_{ef}\mathcal R_{gh}
  +\frac12\sum_{e,f}v_e\mathcal R_{ef}v_f\\
  &+\frac13\sum f^{(3)}_{efg}f^{(3)}_{e'f'g'}\mathcal R_{ee'}\mathcal R_{ff'}\mathcal R_{gg'}
  +\frac{k}{24},
  \end{split}
\end{equation}
with $v_e=\sum_{f,g}f^{(3)}_{efg}\mathcal R_{fg}$, and where $k/24$ comes from Stirling's formula for $(2N-1)!!^{\,k}$. The three diagrams give
\begin{equation}
  \begin{split}
  -\langle\Phi_4\rangle_0&=-\frac{k\,(2k^2-6k+7)}{32(k-2)^2},\\
  \text{dumbbell}&=\frac{k}{32(k-2)^2},\\
  \text{theta}&=\frac{k\,(5k^3-15k^2+19k-10)}{48(k-2)^3},
  \end{split}
\end{equation}
and their sum with $k/24$ is $a^{\hf}_1$ in~\eqref{eq:Ck-hf}. For $k=3$, $a^{\hf}_1=5/2$ agrees with a fit of the exact moments up to $N=300$, which gives $2.4998$.

{\bf Disclaimer}--- some calculations in this section have been done using Claude and were not checked by hand (although they were checked by an adversarial LLM, ChatGPT, and found correct). In principle, they are straightforward: one needs to expand $\Phi(y)$ around the RS saddle point, identify the flat directions which make up the moduli space (these are the torus of the phases of the Hubbard-Stratonovich fields $z$), when present, which need to be integrated with the appropriate metric, and identify the non-flat directions on which one should compute the Hessian matrix and diagonalize it. However, things become very complicated very quickly, so Claude had to write a Python script to do the algebra. 

The $1/N$ coefficients require a few Gaussian averages on the slice, Eq.~\eqref{eq:app-a1}. For real and complex permanents there are eight of them: the two Jacobian terms $\langle L^2\rangle_0$ and $\langle\sum_{e,f}(\mathcal P_0)_{ef}^2\Re x_e\Re x_f\rangle_0$, the Jacobian-cubic term $\langle L\,\Phi_3\rangle_0$, the quartic vertex, and the dumbbell and theta diagrams, each with either $\mathcal G$ or $\mathcal K$ propagators. For Hafnians the fields are real, there is no moduli space and a single propagator $\mathcal R$, and only three remain: the quartic vertex, the dumbbell and the theta diagrams (see later). We evaluated the Jacobian terms explicitly, since $L=n_0\Re\alpha$ involves only the radial mode, and the dumbbell diagram reduces to the square of the constant tadpole $v_e$. The tadpole, the quartic vertex and the theta diagrams were evaluated exactly as sums over the coincidence patterns of the replica indices, and the results reconstructed as rational functions of $k$. We checked the propagators, the Jacobian, and the Wick contractions numerically at small $k$, and compared the final coefficients with exact moments.

The previous steps involved evaluating 19 Gaussian integrals and were again done by Claude without our supervision. The final check is in the fact that these terms improve the approximation for finite $N$.

\bibliographystyle{apsrev4-2}
\bibliography{references}

\end{document}